# Optical Visualization of Carrier Surfing in 2D Monolayers Driven by Surface Acoustic Waves

*Xueqian Sun, Shuyao Qiu, Hao Qin and Yuerui Lu[1,2*]*

Xueqian Sun, Shuyao Qiu, Hao Qin and Yuerui Lu

School of Engineering, College of Engineering, Computing and Cybernetics, The Australian National University, Canberra, ACT, 2601, Australia

Yuerui Lu

Australian Research Council Centre of Excellence for Quantum Computation and Communication Technology, the Australian National University, Canberra, ACT, 2601 Australia

*****E-mail: Yuerui Lu (yuerui.lu@anu.edu.au)

**Abstract:**

Charge carrier transport is pivotal in advancing nanoelectronics. Despite progress in exciton transport within ultra-thin semiconductors, the intertwined transport of free carriers and excitons presents challenges. Surface Acoustic Waves (SAWs) offer a compelling solution, enabling remote, real-time control of excitonic states at room temperature via surfing carriers in 2D materials—a relatively unexplored domain. SAWs create a versatile platform for tailoring excitonic states from microwave to optical frequencies. This study first demonstrates a simple route to visualize directional light transport and carriers drift driven by non-perfect Rayleigh-SAWs. We observed a maximum drift velocity of ~16.4 μm/s for ionized carriers with SAW, significantly surpassing their natural movement in monolayers, though free electrons drift remains in the order of ~$10^3$ m/s. Enhanced exciton emission was achieved through standing SAWs, generating periodic oscillations. By combining traveling and standing wave portions, controllable on-demand single-chip emission is feasible. Our findings open avenues for light manipulation, photonic circuits, and on-chip communications technologies.

## Introduction

The vast technological potential of atomically thin 2D materials, particularly the prominent family of transition metal dichalcogenides (TMDs), has gained significant attention over the past decade. This is largely due to their tightly bound excitons, resulting from strong quantum confinement and reduced dielectric screening, which possess much stronger binding energies compared to excitons in typical III-V semiconductors like GaAs quantum wells[1], thereby underscoring their practical applications at room temperature. The ultra-thin nature of 2D materials makes them highly susceptible to external environmental stimuli[2, 3], enabling flexible manipulation of their optical response through various means, such as mechanical strains[4, 5] and electric fields[6, 7]. Both have been extensively studied as strategies for tuning emission energies, inducing band gap transitions, dissociating excitons, and facilitating exciton transport. This high degree of tunability further emphasizes their advanced light-matter interactions and potential in diverse applications, including efficient energy conversion[8], sensing[9], light emission and information communication[10]. Surface acoustic waves (SAWs) serve as a nanoscale "earthquake" on chip, offering a compelling platform for simultaneously modulating materials through piezoelectric and strain fields. It provides a powerful, reversible method with several advantages over typical external modulation, including minimal loss, non-invasive, contact-free and damage-free operation, all while preserving the materials' intrinsic optical quality.

Unlike static modulation, this is a dynamic approach to mechanically and physically transferring particles while spatially and temporally modulating and confining them in SAW-induced potential wells. This can be employed to investigate the long-range transfer of light in 2D materials, paving the way for scalable quantum computing. It provides a crucial step toward the development of excitonic devices by addressing the significant challenge posed by the nanosecond-range short radiative lifetime[1, 11] of their excitons, allowing particles to be stored, manipulated and transported for a longer period between different locations[12]. Although the alternative interlayer excitons (IXs) with longer lifetime was proposed to solve the issue and demonstrated long-range transport with quality emission[13, 14], SAW devices illuminate a new pathway toward flying qubits or spins, which is particularly promising for quantum communication[15], further demonstrating that photon coherence can be preserved over a long distance. The latest research on SAW modulation of TMD IXs has laid a solid foundation for flying photonic qubits[16], and the funnelling induced by traveling strain profiles has proven to be an efficient conveyor[17] for monolayer excitons. Additionally, SAWs have a long-standing history of having a profound influence on semiconductor systems[18-22] and have been studied for more than a decade. However, the mechanisms of carrier transportation and how mobile carriers dynamically affect materials remain unrevealed. The transport and far-recombination of mobile carriers have never been experimentally imaged in monolayers, leaving a lack of direct evidence for effective propagation at room temperature.

Although excitons dominate the emission in TMD materials, charge carriers, particularly in transport studies, cannot be overlooked. Free carriers, especially electrons, are expected to drift along with the traveling waves at unconventional velocities. Tackling their transport is often complicated due to their general interconnection with excitonic transport[23]. In this study, we optically visualized the propagation of carriers with a drift velocity of approximately 16.4 μm/s, comparable to theoretical predictions for electric-field-dominated electron transport in a wire[24] (Supplementary Text 5) and at least 40 times faster than their free movement in TMDs, by leveraging the piezoelectric filed-dominated SAW effect in intrinsically N-doped 1L $WS_2$[25]. We reveal a leaking property along the propagation, leading to a non-perfect Rayleigh wave even though SAWs are not strictly classified as evanescent waves[26]. A moderate-frequency SAW device (~178 MHZ, Fig. S1) was involved to ensure good coupling efficiency of the applied RF power to the resonator[20]. The transported ionized carriers recombine at a distant location, achieving remote lasing. We further tuned the SAW driving power and excitation pumping power to emphasize the importance of threshold excitation conditions for efficient modulation. With the strategic optimization of SAW device and the addition of another actuator with aligned frequency and amplitude, we demonstrated distinct modulation under a standing wave. This unlocks advanced emission engineering with a high level of tunability through acoustic wave interference, adjustable to combine traveling and standing wave modes. Our work paves the way for multifunctional, scalable technologies in modern microelectronics, such as computer chips and optoelectronic circuits, driving the development of next-generation devices with advanced photonic information processing capabilities.

## Results and discussion

### Acoustically driven optical modulation in TMD Monolayers

SAWs serve as a versatile tool, offering a hybridized manipulation that enables good control over the modulation of target materials[16, 17, 19, 21, 27]. Figure 1a illustrates the experimental configuration and device geometry designed in this study, where a monolayer $WS_2$ sample was transferred adjacent to the output port of the IDTs and positioned at the centre of the generated SAW beam. The waves were launched in the direction indicated by the arrow, powered by RF signals and travelled toward the sample. An elongated $WS_2$ sample was selected to explore the SAW effects at various positions along the wave and due to its higher carrier mobility among the commonly studied group-6 TMD monolayers[28]. The sample was aligned with the propagation direction of SAW, as shown in the optical image in Figure 1b. The zoomed-in PL mapping reveals the relatively spatially uniform quality of the sample across the entire region. The typical manipulation created by SAW can be classified into two categories, each leading to lattice deformations that result in either antisymmetric (Type I) or symmetric (Type II) band diagram modulations.

These two types of modulation intervene simultaneously and periodically on the materials but are driven by strain and piezoelectric fields, respectively. Type-I modulation induces spatiotemporal traps with varying band gaps (Fig. S2), where the minimum occurs in regions of maximum tension and the maximum in regions of maximum compression[29]. Type-II, on the contrary, governed by the lateral piezoelectric field, spatially separates free electrons and holes by ionizing excitons and creating piezoelectric potential wells, maintaining a constant band gap[19] (Fig. 1c). These wells facilitate charge carrier transport, where electrons and holes occupy opposite potential pockets along the wave[22]. Interestingly, the intrinsic free electrons in typically N-doped $WS_2$ also propagate along the wave, and both dissociated carriers and intrinsic electrons actually move with their velocities determined by their coupling strength with the wave (discussed later). Both effects can collectively act on the materials, but the modulation from the piezoelectric field is typically two orders of magnitude stronger than that from the strain field[16, 27]. This should be particularly evident in our case, where the involvement of monolayer TMDs with strong in-plane piezoelectricity leads to a more dominant Type-II modulation[30].

Directly placing $WS_2$ on a $LiNbO_3$ substrate with a high dielectric constant[31] allowed for efficient exciton dissociation, as demonstrated by the strong PL quenching, near absence of emission, along the SAW propagation path (Fig. 1d). This separation of free electrons and holes in real space increases their lifetime while it inhibits their radiative recombination[32]. Additionally, the acoustically transported charge carriers were moved out of the detection region, further reducing the observed local photoluminescence and validating the surfing of carriers. As the detection spot moved farther from the IDTs, the quenching effect was significantly reduced and eventually reversed (Fig. S4), enhancing the PL emission of $WS_2$ at the sample's edge. This enhancement stems from the increased radiative recombination of accumulated dissociated free carriers at this location, as the propagating carriers are unable to travel further due to the sample size limitation. It may also arise from the enhanced neutralization of the local $WS_2$ region because of the equal number of electrons and holes being transferred from the other side.

To quantify the SAW-induced quenching, we introduced a parameter, $\eta = I_{WS_2\,saw\,off} / I_{WS_2\,SAW\,on}$, representing the PL intensity ratio without and with SAW modulation. We uniformly divided the sample and observed that η values were considerably higher at locations closer to the IDTs (Fig. 1e, top), declining almost linearly with distance from the transducer. The accompanying redshift in exciton emission further confirms the dominance of Type-II modulation, triggered by a Stark effect[19, 33] from the propagating electric field, and follows a similar decreasing trend (Fig. 1e, bottom). This suggests a gradual attenuation of SAW strength along the propagation path, despite theoretical predictions that Rayleigh waves maintain amplitude across the surface[34, 35]. The specific orientation of $LiNbO_3$ (128° Y-cut, Supplementary Text 3) used in our measurements was found to generate traditional Rayleigh-type SAWs[36]. However, as the wave propagates across the substrate surface and interacts with 2D materials, it undergoes mode conversion, leading to the formation of leaky SAWs (Fig. 1f), resembling SAW behavior in fluid media[37, 38]. The extracted ratio of trion to exciton and time-resolved photoluminescence (TRPL) (Fig. S5 & 6 and Supplementary Text 3 & 4) as a function of spots provide additional insights to support our observations.

**Visualized surface-acoustic-wave induced engineering and carriers transport in 2D semiconductors**

To further clarify our interpretation of the experimental results, spatially temporal resolved PL mapping, rather than steady-state PL spectra, was conducted to visually investigate the transient response of the material's photonic tunability with drifting carriers under the influence of SAW. The measurement setup schematic in Figure 2a shows uniform illumination across the sample, leading to continuous exciton excitation throughout the surface. The whole process of dynamic SAW modulation on the monolayer emission was recorded in Movie S1, with key frames highlighted in Figures 2b and 2c, monitoring the propagation of carriers and revealing two distinct rounds of in-plane transitions. The entire procedure can be divided into six key steps, beginning with the inherent state of the materials itself, "SAW off" (before modulation), followed by the instantaneous response to SAW, "SAW on 0s", and the stable phase with ongoing SAW modulation, "SAW on 1.5s". The next steps are the immediate response after removing SAW, "SAW off 0s", the intermediate state during recovery, "SAW off 5s", and finally, the state after the material has returned to its initial condition, "SAW off 60s". Figure 3a and 3b represent the schematic diagrams corresponding to these states.

Specifically, when the SAW is first applied to the materials, the spot closest to the IDTs becomes the brightest (Fig. 2b(i) and 3a(i)) with a rapid transition, this brightest spot then moves toward the opposite side, becoming brightest at the edge of the sample (Fig. 2b(ii) and 3a(ii)). The initial brightest spot closest to the IDT is attributed to the greatest neutralization, as the propagating SAW first displaces free electrons away from the right side while dissociating neutral excitons. This drift velocity of free electrons under the piezoelectric field[39, 40] is exceptionally fast (Supplementary Text 5), making it difficult to capture their actual movement. The subsequent quenching from right to left is ascribed to the progressive ionization of excitons, as previously explained, with free carriers traveling along the SAW and eventually recombining at the sample's edge, suggesting potential of SAW devices for remote lasing, as observed in quantum wells[12, 15, 29, 41] and TMD heterostructures[16]. From this transport, we determined the drift velocity of dissociated carriers to be around 16.4 μm/s, several orders of magnitude slower than that of intrinsic free electrons under the same piezoelectric field. This is attributed to the significantly enhanced carrier density, consistent with the transport of electrons in a wire driven by an electric field (Supplementary Text 5). Additionally, we confirmed the free carrier transport and remote emissions through PL spatial mapping under single-spot excitation with reduced exciton density (Fig. S7). It is demonstrated through bright emission at a distant spot and the effective quenching in the local area in our work. This presents a promising platform, analogous to observations in quantum wells, with the potential to enable high-temperature photonic memory devices[12]. It highlights the possibility of controlled, customised on-chip emissions and single emitters[18, 42, 43], particularly when integrated the SAW devices with tailored defects like quantum dots.

After turning off the SAW, the brightest spot began to shift gradually toward the right (closest to IDTs) (Fig. 2c(i) and 3b(i)), driven by the backflow of exciton flux (accumulated excitons at the sample edge where the dissociated electrons and holes recombined to reform excitons) and its intrinsic carriers once the SAW-driven force was removed. They then reached the opposite boundary of the sample and rebounded leftward (Fig. 2c(ii) and 3b(ii)). Eventually, sample resumed with their neutral drifting after a longer period (Fig. S8 and Supplementary Text 6), with an even distribution of free carriers, restoring the emission state to its initial condition (Fig. 2c(iii) and 3b(iii)). Based on this, we estimated the neutral drift velocity of carriers to be approximately 0.41 μm/s, at least 40 times slower than that with the SAW driving force. The ability to capture all these steps with a CMOS sensor indicates weak coupling between the surfing carriers and the propagating wave, where the drift velocity of carriers is insufficient to keep pace with the traveling SAW[44], causing them to move oppositely with SAW at a lagging speed (black arrows in Fig. S2).

Figure 3c and 3d demonstrate the further tunability of SAW effects with varying SAW driving power and laser excitation power for both quenching and enhancing areas. The quenching and enhancing factors (Supplementary Text 7) exhibit an increasing trend as SAW modulation strengthens with higher RF power, while the SAW effects show the opposite behavior with increasing pumping power, consistent with previous reports[17, 21, 30]. Intriguingly, before the quenching and enhancing factors decreased with higher excitation density, the SAW effects on both regions in our study were initially increased (orange dashed area). This can be explained by a more prominent SAW effect due to the rising exciton population. However, when the excitation power is further increased beyond a certain threshold, the piezoelectric field becomes screened by a large number of optically generated free carriers[12, 17]. Even if the piezoelectric field is not fully screened, the SAW modulation becomes less effective due to the overwhelming presence of carriers, making it difficult for the traveling wave to displace them (Fig. S9).

**Acoustic-excitonic manipulation accessed by regulated carriers with standing waves**

When another set of IDTs is introduced at the opposite end with the same resonant frequency ($f_{SAW1} = f_{SAW2}$), two acoustic waves are launched by actuating the pair of IDTs and travel toward each other with the same pace. A new collective wave is created when these waves collide and interfere, giving rise to a synergistic effect on the material. This modulation produces a periodic oscillation in both time and space, distinct from that of a traveling wave. Particularly, when both SAWs are precisely actuated with equal amplitude (Fig. 4a), the superposition of two opposing but identical waves leads to constructive and destructive interference, forming a standing wave[12, 45, 46] (Fig. 4b). This creates alternating regions of pressure nodes and anti-nodes, corresponding to varying strain through compression and tension in the material, resembling a type-I modulation. These pressure fluctuations generate varying acoustic forces that are expected to alternatively influence the PL emission of 2D materials and both compression and tension strain have been reported to effectively affect the PL emission of TMD monolayers[47-49]. In our study, both X-axis standing SAWs (Fig. 4c) and Y-axis standing SAWs (Fig. S10) applied parallel and perpendicular to the sample orientation, showcase a strong enhancement in PL intensity at all the spots compared to the unmodulated $WS_2$ sample (Fig. S12), with comparable strength, as shown in Figure 4d. This underscores the efficient orientation-independent modulation from SAWs and reveals a significant difference from the modulation induced by traveling SAWs and strain studies generated by wrinkles, folding, or bending[48-50]. This enhancement is attributed to the improved quantum yield by small strain-suppressed exciton-exciton annihilation (Supplementary Text 8)[51, 52]. The increased overlap in time and space of the electron and hole wavefunctions trapped by the two incoming SAW potentials (Fig. S11) also increases the possibility of radiative recombination, creating enhanced photon emission compared to the modulated emission with a propagating SAW[12]. Importantly, while all spots along the sample give rise to an enhancement upon the application of standing SAWs, the enhancement factor demonstrates spatially periodic fluctuations corresponding to the acoustic wavelength (Fig. 4e, Movie S2 and Supplementary Text 9). Similar to the propagating wave discussed earlier regarding the power-dependent SAW effect, it suggests improved modulation by harnessing the driving force from the RF signal generator (Fig. 4f). The analogous saturation behavior in enhancement at higher power may be attributed to the full filling of the potential pockets generated by SAW, consistent with the phenomenon observed with travelling SAWs. By meticulously controlling the amplitude of opposing waves, the desired controllable emission can be realized by leveraging the distinct modulation effects from the combination[45] of standing and traveling wave components or localized purely traveling wave characteristics.

## Conclusion

This study demonstrates the remarkable potential of SAW modulation in 2D semiconductors for creating on-demand emitters capable of light storage by efficiently trapping electrons and holes separately in moving SAW potential wells. It enables the generation of high-temperature remote lasers with precise, contact-free control over time, space, and array design. We first optically visualized the surfing of carriers through the efficient dissociation of neutral excitons by non-perfect Rayleigh-type SAWs, driven primarily by strong piezoelectric fields, revealing distinct drift velocities between ionized and intrinsic free carriers under the same SAW conditions. Experimentally, we extracted a notably enhanced drift of free carriers under SAW modulation compared to their natural movement without an external driving force. This highlights the weak coupling between free carriers and propagating acoustic waves and underscores the critical role of SAW in modulating light emission via carrier transport, with further tunability through the driving force and excitation power. In standing acoustic waves, we observed an exceptional enhancement in exciton emission, attributed to the improved quantum yield from the application of slight mechanical strain. This direct optical visualization of surfing carriers and the resulting reversible modulation of exciton emission offer insights into the non-invasive, dynamic manipulation enabled by SAWs. Promising future work includes integrating SAW devices with engineered defects in monolayer semiconductors to achieve room-temperature single-photon emitters, benefiting quantum computation and information processing. Another direction is extending SAW modulation to heterobilayers with spatially indirect excitons, which have longer lifetimes, allowing greater flexibility for manipulation and storage. Additionally, SAWs offer exciting opportunities in the emerging field of moiré potentials. Investigating the interplay between SAWs and moiré-confined excitons is expected to enable tunable quantum phenomena, expanding SAW-enabled optoelectronic applications. This study lays the foundation for these advancements, establishing SAW-based platforms as key tools for advanced light control and next-generation photonic technologies.

**Experimental Methods**

*Device Fabrication and Characterization*: $WS_2$ monolayers were mechanically exfoliated from bulk 2H crystals (purchased from HQ Graphene) and selected for their elongated shape under a microscope. These flakes were identified by their optical contrast and further confirmed using photoluminescence and Raman spectroscopy. The monolayer quality was also assessed and verified using atomic force microscopy (AFM) and Phase shifting interferometry (PSI)[53] (Fig. S3). The monolayers were then dry transferred onto commercially available 128° Y-cut $LiNbO_3$ wafers (purchased from OST Photonics), which were pre-patterned with Interdigital transducers (IDTs) using standard photolithography technique followed by a metal lift-off process. To ensure efficient acoustic wave propagation, the fabrication of the SAW device began with cleaning the substrate to remove surface contaminants. A photoresist layer was then uniformly spin-coated onto the substrate and exposed to UV light through a photolithographic mask to define the IDT structure. After developing the photoresist, a thin metal layer (aluminium or gold) was deposited using electron beam evaporation. Finally, the unwanted metal was removed through the lift-off process, leaving behind well-defined IDT electrodes. The final SAW device was inspected under an optical microscope to confirm pattern accuracy and uniformity. All fabrication steps were carried out in a cleanroom environment to ensure high device quality and consistency. The IDT electrodes in this study consist of 25 pairs of fingers, with a 10 nm thick chromium adhesive layer and 100 nm thick aluminium (or gold), designed for an acoustic wavelength of $\lambda_{SAW}$ = 22 µm with a corresponding SAW frequency of $f_{SAW}$ = 178 MHz. The $WS_2$ monolayers were positioned in the middle of the SAW beam, close to the output port of the IDT, along the delay line which is defined along the in-plane X direction of the $LiNbO_3$ piezoelectric substrates.

*Measurement of Acoustic response from the designed SAW devices*: The acoustic response of SAW devices was characterized using a vector network analyzer (R&S ZNA43-4ports) complemented by an S-parameter test unit. The S11 results of the acoustic response from SAW devices can be found in the Supplementary materials, Figure 1.

*Optical Characterization under SAW excitation*: Micro-PL measurements were conducted using a Horiba LabRAM system equipped with a confocal microscope, a charge-coupled device (CCD) Si detector (detection range between 400 and 1000 nm), and a 532 nm diode-pumped solid-state (DPSS) laser as the excitation source. The laser light was focused on the sample surface via a 50x objective lens (numerical aperture = 0.55), and the laser beam was gaussian in nature. PL mapping was carried out on a custom-built fluorescence microscope with an 50x Nikon objective lens and a CMOS (Complementary Metal Oxide Semiconductor) image sensor with 500 nm short pass and 550 nm long pass filters. The SAW devices were excited at the resonance frequencies *via* a radio frequency power generator connected to the microscope-compatible chamber with micro electrical probes. The device was placed into the chamber and all PL spectra were captured when the surface of the device reached a thermal equilibrium after applying SAW.

Time-resolved PL (TRPL) measurements were carried out in a setup that incorporates micro-PL spectroscopy with a time-correlated single-photon counting (TCSPC) system. A linearly polarized pulsed laser (frequency doubled to 522 nm, with a 300 fs pulse width) was directed to a high numerical aperture (NA = 0.7) objective (Nikon S Plan 60×). The PL signal was collected by a grating spectrometer, thereby either recording the PL spectrum through the CCD (Princeton Instruments, PIXIS) or detecting the PL intensity decay by a Si single-photon avalanche diode and the TCSPC (PicoHarp 300) system. All the PL spectra were corrected for the instrument response. We fixed the spectrometer at specific exciton resonances to collect the photons at respective wavelengths for the TRPL measurements. All the measured decay curves were deconvoluted with respect to the instrument response function (IRF) and then were fitted using the following equation: $I = A\,exp(-\frac{t}{\tau_1}) + B\,exp(-\frac{t}{\tau_2}) + C$, where I is the PL intensity, A, B, and C are constants, ‘t’ is time, and $\tau_1$ is faster decay rate and $\tau_2$ is the slower decay rate, indicating emission lifetimes for different decay processes. The short lifetime is effectively considered to represent non-radiative lifetime decay while the longer component is indicative of the radiative recombination time, which is what we focused. The weighted mean lifetime was calculated using the area under the curve of the PL lifetime decay curve. All optical spectroscopy measurements were carried out at room temperature.

**Supporting Information**

Supporting Information is available from the Wiley Online Library or from the author.

**Acknowledgements**

The authors acknowledge funding support from Australian Research Council (grant No: DP220102219, DP180103238, LE200100032), ARC Centre of Excellence in Quantum Computation and Communication Technology (project number CE170100012) and the National Health and Medical Research Council (NHMRC ID: GA275784). The authors would also like to thank Professor Chennupati Jagadish, Professor Daniel MacDonald and Dr Hieu Nguyen from the Australian National University for their facility support.

**Author Contributions**

Y. L. supervised this study; Y. L. and X. S. conceived this work and X. S. fabricated all the samples used in this study; X. S. performed the optical experiments; X. S. and Y. L. analyzed the data, developed the models and interpreted the results; X. S. processed experimental raw data and created all the figure plots; S. Q. contributed schematic figures; X. S. drafted the manuscript with guidance from Y. L.; and all authors participated in manuscript editing and discussion.

**Conflict of Interest**

The authors declare no conflict of interest.

**Data Availability Statement**

The data that support the findings of this study are available from the corresponding author upon reasonable request.

## References

1. Sun, X., E. Malic, and Y. Lu, *Dipolar many-body complexes and their interactions in stacked 2D heterobilayers.* Nature Reviews Physics, 2024: p. 1-16.
2. Zhang, Q., J. Zhang, S. Wan, W. Wang, and L. Fu, *Stimuli-responsive 2D materials beyond graphene.* Advanced Functional Materials, 2018. **28**(45): p. 1802500.
3. Glavin, N.R., R. Rao, V. Varshney, E. Bianco, A. Apte, A. Roy, E. Ringe, and P.M. Ajayan, *Emerging applications of elemental 2D materials.* Advanced Materials, 2020. **32**(7): p. 1904302.
4. Gelly, R.J., D. Renaud, X. Liao, B. Pingault, S. Bogdanovic, G. Scuri, K. Watanabe, T. Taniguchi, B. Urbaszek, and H. Park, *Probing dark exciton navigation through a local strain landscape in a WSe2 monolayer.* Nature Communications, 2022. **13**(1): p. 232.
5. Ripin, A., R. Peng, X. Zhang, S. Chakravarthi, M. He, X. Xu, K.-M. Fu, T. Cao, and M. Li, *Tunable phononic coupling in excitonic quantum emitters.* Nature Nanotechnology, 2023. **18**(9): p. 1020-1026.
6. Chakraborty, B., J. Gu, Z. Sun, M. Khatoniar, R. Bushati, A.L. Boehmke, R. Koots, and V.M. Menon, *Control of strong light–matter interaction in monolayer WS2 through electric field gating.* Nano letters, 2018. **18**(10): p. 6455-6460.
7. Massicotte, M., F. Vialla, P. Schmidt, M.B. Lundeberg, S. Latini, S. Haastrup, M. Danovich, D. Davydovskaya, K. Watanabe, and T. Taniguchi, *Dissociation of two-dimensional excitons in monolayer WSe2.* Nature communications, 2018. **9**(1): p. 1633.
8. Tao, H., Q. Fan, T. Ma, S. Liu, H. Gysling, J. Texter, F. Guo, and Z. Sun, *Two-dimensional materials for energy conversion and storage.* Progress in Materials Science, 2020. **111**: p. 100637.
9. Li, B.L., J. Wang, H.L. Zou, S. Garaj, C.T. Lim, J. Xie, N.B. Li, and D.T. Leong, *Low-dimensional transition metal dichalcogenide nanostructures based sensors.* Advanced Functional Materials, 2016. **26**(39): p. 7034-7056.
10. Butov, L.V., *Excitonic devices.* Superlattices and Microstructures, 2017. **108**: p. 2-26.
11. Sun, X., Z. Lu, and Y. Lu, *Enhanced interactions of excitonic complexes in free-standing WS2.* Nanoscale, 2023. **15**(48): p. 19533-19545.
12. Rocke, C., S. Zimmermann, A. Wixforth, J.P. Kotthaus, G. Böhm, and G. Weimann, *Acoustically driven storage of light in a quantum well.* Physical Review Letters, 1997. **78**(21): p. 4099.
13. Lee, H., Y.B. Kim, J.W. Ryu, S. Kim, J. Bae, Y. Koo, D. Jang, and K.-D. Park, *Recent progress of exciton transport in two-dimensional semiconductors.* Nano Convergence, 2023. **10**(1): p. 57.
14. Unuchek, D., A. Ciarrocchi, A. Avsar, K. Watanabe, T. Taniguchi, and A. Kis, *Room-temperature electrical control of exciton flux in a van der Waals heterostructure.* Nature, 2018. **560**(7718): p. 340-344.
15. Helgers, P.L.J., J.A.H. Stotz, H. Sanada, Y. Kunihashi, K. Biermann, and P.V. Santos, *Flying electron spin control gates.* Nature Communications, 2022. **13**(1): p. 5384.
16. Peng, R., A. Ripin, Y. Ye, J. Zhu, C. Wu, S. Lee, H. Li, T. Taniguchi, K. Watanabe, T. Cao, X. Xu, and M. Li, *Long-range transport of 2D excitons with acoustic waves.* Nature Communications, 2022. **13**(1): p. 1334.
17. Datta, K., Z. Lyu, Z. Li, T. Taniguchi, K. Watanabe, and P.B. Deotare, *Spatiotemporally controlled room-temperature exciton transport under dynamic strain.* Nature Photonics, 2022. **16**(3): p. 242-247.

18. Couto Jr, O., S. Lazić, F. Iikawa, J. Stotz, U. Jahn, R. Hey, and P. Santos, *Photon anti-bunching in acoustically pumped quantum dots.* Nature Photonics, 2009. **3**(11): p. 645-648.
19. Scolfaro, D., M. Finamor, L.O. Trinchão, B.L.T. Rosa, A. Chaves, P.V. Santos, F. Iikawa, and O.D.D. Couto, Jr., *Acoustically Driven Stark Effect in Transition Metal Dichalcogenide Monolayers.* ACS Nano, 2021. **15**(9): p. 15371-15380.
20. Kinzel, J.B., D. Rudolph, M. Bichler, G. Abstreiter, J.J. Finley, G. Koblmüller, A. Wixforth, and H.J. Krenner, *Directional and Dynamic Modulation of the Optical Emission of an Individual GaAs Nanowire Using Surface Acoustic Waves.* Nano Letters, 2011. **11**(4): p. 1512-1517.
21. Datta, K., Z. Li, Z. Lyu, and P.B. Deotare, *Piezoelectric Modulation of Excitonic Properties in Monolayer WSe2 under Strong Dielectric Screening.* ACS Nano, 2021. **15**(7): p. 12334-12341.
22. Weinreich, G., *Acoustodynamic Effects in Semiconductors.* Physical Review, 1956. **104**(2): p. 321-324.
23. Rosati, R., I. Paradisanos, E. Malic, and B. Urbaszek, *Two dimensional semiconductors: optical and electronic properties.* arXiv preprint arXiv:2405.04222, 2024.
24. *Model of Conduction in Metals.*
25. Zinkiewicz, M., T. Wozniak, T. Kazimierczuk, P. Kapuscinski, K. Oreszczuk, M. Grzeszczyk, M. Bartoš, K. Nogajewski, K. Watanabe, and T. Taniguchi, *Excitonic complexes in n-doped WS2 monolayer.* Nano letters, 2021. **21**(6): p. 2519-2525.
26. Tarasenko, A., R. Čtvrtlík, and R. Kudělka, *Theoretical and experimental revision of surface acoustic waves on the (100) plane of silicon.* Scientific Reports, 2021. **11**(1): p. 2845.
27. Fandan, R., J. Pedrós, and F. Calle, *Exciton–Plasmon Coupling in 2D Semiconductors Accessed by Surface Acoustic Waves.* ACS Photonics, 2021. **8**(6): p. 1698-1704.
28. Mir, S.H., V.K. Yadav, and J.K. Singh, *Recent Advances in the Carrier Mobility of Two-Dimensional Materials: A Theoretical Perspective.* ACS Omega, 2020. **5**(24): p. 14203-14211.
29. Lazić, S., A. Violante, K. Cohen, R. Hey, R. Rapaport, and P.V. Santos, *Scalable interconnections for remote indirect exciton systems based on acoustic transport.* Physical Review B, 2014. **89**(8): p. 085313.
30. Sheng, L., G. Tai, Y. Yin, C. Hou, and Z. Wu, *Layer-Dependent Exciton Modulation Characteristics of 2D MoS2 Driven by Acoustic Waves.* Advanced Optical Materials, 2021. **9**(3): p. 2001349.
31. Kushibiki, J.-i., I. Takanaga, M. Arakawa, and T. Sannomiya, *Accurate measurements of the acoustical physical constants of LiNbO/sub 3/and LiTaO/sub 3/single crystals.* IEEE transactions on ultrasonics, ferroelectrics, and frequency control, 1999. **46**(5): p. 1315-1323.
32. Huang, T., P. Han, X. Wang, J. Ye, W. Sun, S. Feng, and Y. Zhang, *Theoretical study on dynamic acoustic modulation of free carriers, excitons, and trions in 2D MoS2 flake.* Journal of Physics D: Applied Physics, 2017. **50**(11): p. 114005.
33. Choquer, M., M. Weiß, E.D. Nysten, M. Lienhart, P. Machnikowski, D. Wigger, H.J. Krenner, and G. Moody, *Quantum control of optically active artificial atoms with surface acoustic waves.* IEEE Transactions on Quantum Engineering, 2022. **3**: p. 1-17.
34. Caliendo, C., *Acoustoelectric effect for rayleigh wave in ZnO produced by an inhomogeneous in-depth electrical conductivity profile.* Sensors, 2023. **23**(3): p. 1422.

35. Dransfeld, K. and E. Salzmann, *Excitation, detection, and attenuation of high-frequency elastic surface waves.* Physical acoustics, 2012. **7**: p. 219-272.
36. Nie, X., X. Wu, Y. Wang, S. Ban, Z. Lei, J. Yi, Y. Liu, and Y. Liu, *Surface acoustic wave induced phenomena in two-dimensional materials.* Nanoscale Horizons, 2023. **8**(2): p. 158-175.
37. Li, S. and V.R. Bhethanabotla, *Design of a Portable Orthogonal Surface Acoustic Wave Sensor System for Simultaneous Sensing and Removal of Nonspecifically Bound Proteins.* Sensors, 2019. **19**(18): p. 3876.
38. Zhang, Y. and X. Chen, *Particle separation in microfluidics using different modal ultrasonic standing waves.* Ultrasonics Sonochemistry, 2021. **75**: p. 105603.
39. Jin, Z., X. Li, J.T. Mullen, and K.W. Kim, *Intrinsic transport properties of electrons and holes in monolayer transition-metal dichalcogenides.* Physical Review B, 2014. **90**(4): p. 045422.
40. Li, X., J.T. Mullen, Z. Jin, K.M. Borysenko, M. Buongiorno Nardelli, and K.W. Kim, *Intrinsic electrical transport properties of monolayer silicene and MoS 2 from first principles.* Physical Review B—Condensed Matter and Materials Physics, 2013. **87**(11): p. 115418.
41. Sogawa, T., P. Santos, S. Zhang, S. Eshlaghi, A. Wieck, and K. Ploog, *Transport and lifetime enhancement of photoexcited spins in GaAs by surface acoustic waves.* Physical review letters, 2001. **87**(27): p. 276601.
42. Yuan, M., K. Biermann, S. Takada, C. Bauerle, and P.V. Santos, *Remotely pumped GHz antibunched emission from single exciton centers in GaAs.* ACS photonics, 2021. **8**(3): p. 758-764.
43. Fuhrmann, D.A., S.M. Thon, H. Kim, D. Bouwmeester, P.M. Petroff, A. Wixforth, and H.J. Krenner, *Dynamic modulation of photonic crystal nanocavities using gigahertz acoustic phonons.* Nature Photonics, 2011. **5**(10): p. 605-609.
44. Violante, A., K. Cohen, S. Lazić, R. Hey, R. Rapaport, and P.V. Santos, *Dynamics of indirect exciton transport by moving acoustic fields.* New Journal of Physics, 2014. **16**(3): p. 033035.
45. Weser, R., A. Winkler, M. Weihnacht, S. Menzel, and H. Schmidt, *The complexity of surface acoustic wave fields used for microfluidic applications.* Ultrasonics, 2020. **106**: p. 106160.
46. Shi, J., H. Huang, Z. Stratton, Y. Huang, and T.J. Huang, *Continuous particle separation in a microfluidic channel via standing surface acoustic waves (SSAW).* Lab on a Chip, 2009. **9**(23): p. 3354-3359.
47. Peng, Z., X. Chen, Y. Fan, D.J. Srolovitz, and D. Lei, *Strain engineering of 2D semiconductors and graphene: from strain fields to band-structure tuning and photonic applications.* Light: Science & Applications, 2020. **9**(1): p. 190.
48. Li, Z., Y. Lv, L. Ren, J. Li, L. Kong, Y. Zeng, Q. Tao, R. Wu, H. Ma, B. Zhao, D. Wang, W. Dang, K. Chen, L. Liao, X. Duan, X. Duan, and Y. Liu, *Efficient strain modulation of 2D materials via polymer encapsulation.* Nature Communications, 2020. **11**(1): p. 1151.
49. Conley, H.J., B. Wang, J.I. Ziegler, R.F. Haglund, Jr., S.T. Pantelides, and K.I. Bolotin, *Bandgap Engineering of Strained Monolayer and Bilayer MoS2.* Nano Letters, 2013. **13**(8): p. 3626-3630.
50. Wang, J., M. Han, Q. Wang, Y. Ji, X. Zhang, R. Shi, Z. Wu, L. Zhang, A. Amini, L. Guo, N. Wang, J. Lin, and C. Cheng, *Strained Epitaxy of Monolayer Transition Metal Dichalcogenides for Wrinkle Arrays.* ACS Nano, 2021. **15**(4): p. 6633-6644.

51. Kim, H., S.Z. Uddin, N. Higashitarumizu, E. Rabani, and A. Javey, *Inhibited nonradiative decay at all exciton densities in monolayer semiconductors.* Science, 2021. **373**(6553): p. 448-452.
52. Lee, Y., J.D.a.S. Forte, A. Chaves, A. Kumar, T.T. Tran, Y. Kim, S. Roy, T. Taniguchi, K. Watanabe, A. Chernikov, J.I. Jang, T. Low, and J. Kim, *Boosting quantum yields in two-dimensional semiconductors via proximal metal plates.* Nature Communications, 2021. **12**(1): p. 7095.
53. Yang, J., Z. Wang, F. Wang, R. Xu, J. Tao, S. Zhang, Q. Qin, B. Luther-Davies, C. Jagadish, Z. Yu, and Y. Lu, *Atomically thin optical lenses and gratings.* Light: Science & Applications, 2016. **5**(3): p. e16046-e16046.

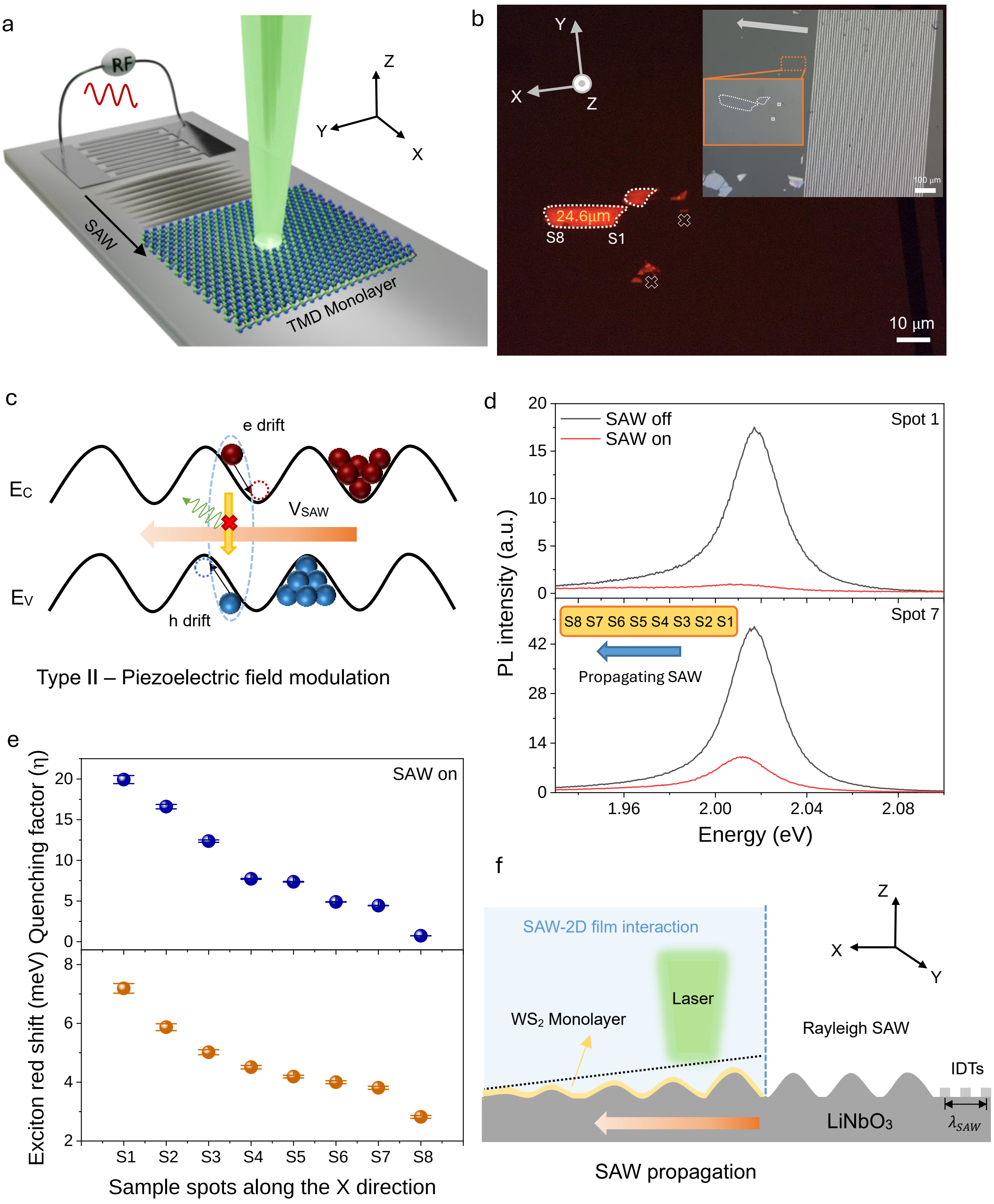
a
RF
SAW
TMD Monolayer
Z
Y
X
b
Y
X
Z
24.6μm
S8
S1
100 μm
10 μm
c
e drift
Ec
VSAW
Ev
h drift
Type II – Piezoelectric field modulation
d
SAW off
SAW on
Spot 1
Spot 7
S8 S7 S6 S5 S4 S3 S2 S1
Propagating SAW
PL intensity (a.u.)
Energy (eV)
e
SAW on
Quenching factor (η)
Exciton red shift (meV)
S1 S2 S3 S4 S5 S6 S7 S8
Sample spots along the X direction
f
SAW-2D film interaction
Laser
WS2 Monolayer
Rayleigh SAW
IDTs
LiNbO3
λSAW
SAW propagation

Figure 1

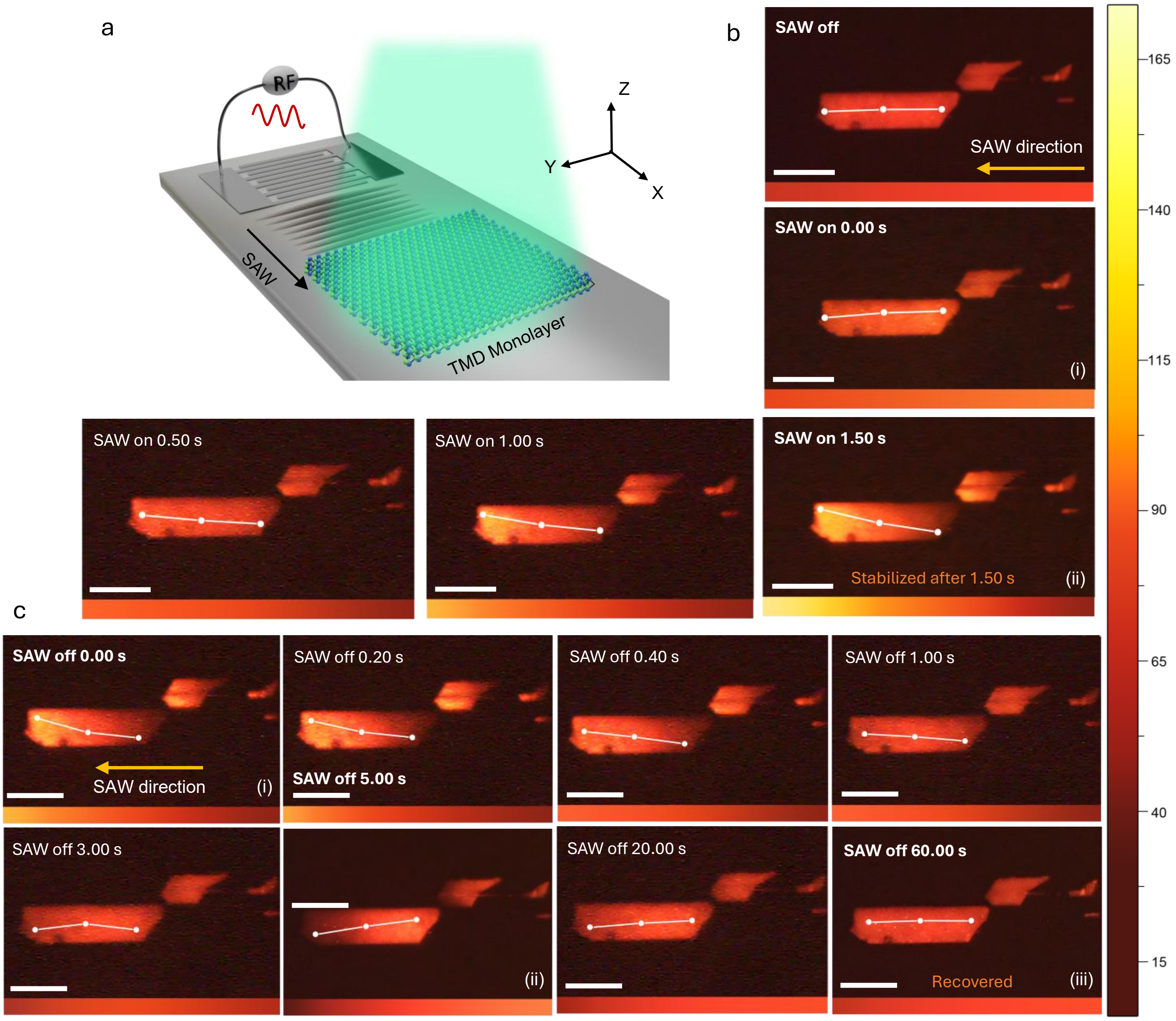

a
RF
Z
Y
X
SAW
TMD Monolayer
b
SAW off
SAW direction
SAW on 0.00 s
(i)
SAW on 0.50 s
SAW on 1.00 s
SAW on 1.50 s
Stabilized after 1.50 s
(ii)
c
SAW off 0.00 s
SAW direction
(i)
SAW off 0.20 s
SAW off 5.00 s
SAW off 0.40 s
SAW off 1.00 s
SAW off 3.00 s
(ii)
SAW off 20.00 s
SAW off 60.00 s
Recovered
(iii)
165
140
115
90
65
40
15

Figure 2

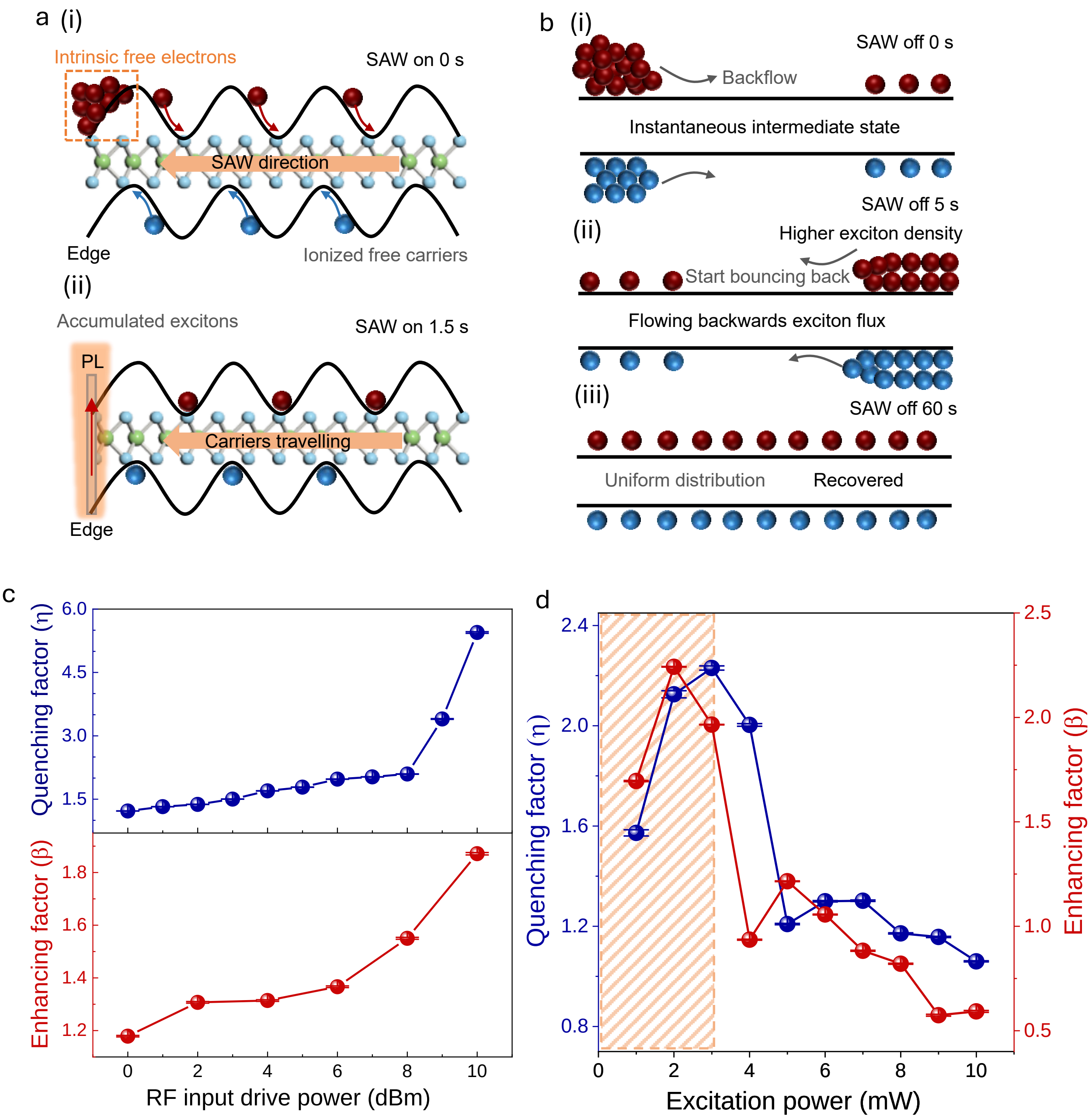

a (i)
Intrinsic free electrons
SAW on 0 s
SAW direction
Edge
Ionized free carriers
(ii)
Accumulated excitons
SAW on 1.5 s
PL
Carriers travelling
Edge
b (i)
SAW off 0 s
Backflow
Instantaneous intermediate state
SAW off 5 s
(ii)
Higher exciton density
Start bouncing back
Flowing backwards exciton flux
(iii)
SAW off 60 s
Uniform distribution
Recovered
c
Quenching factor (η)
Enhancing factor (β)
RF input drive power (dBm)
d
Quenching factor (η)
Enhancing factor (β)
Excitation power (mW)

Figure 3

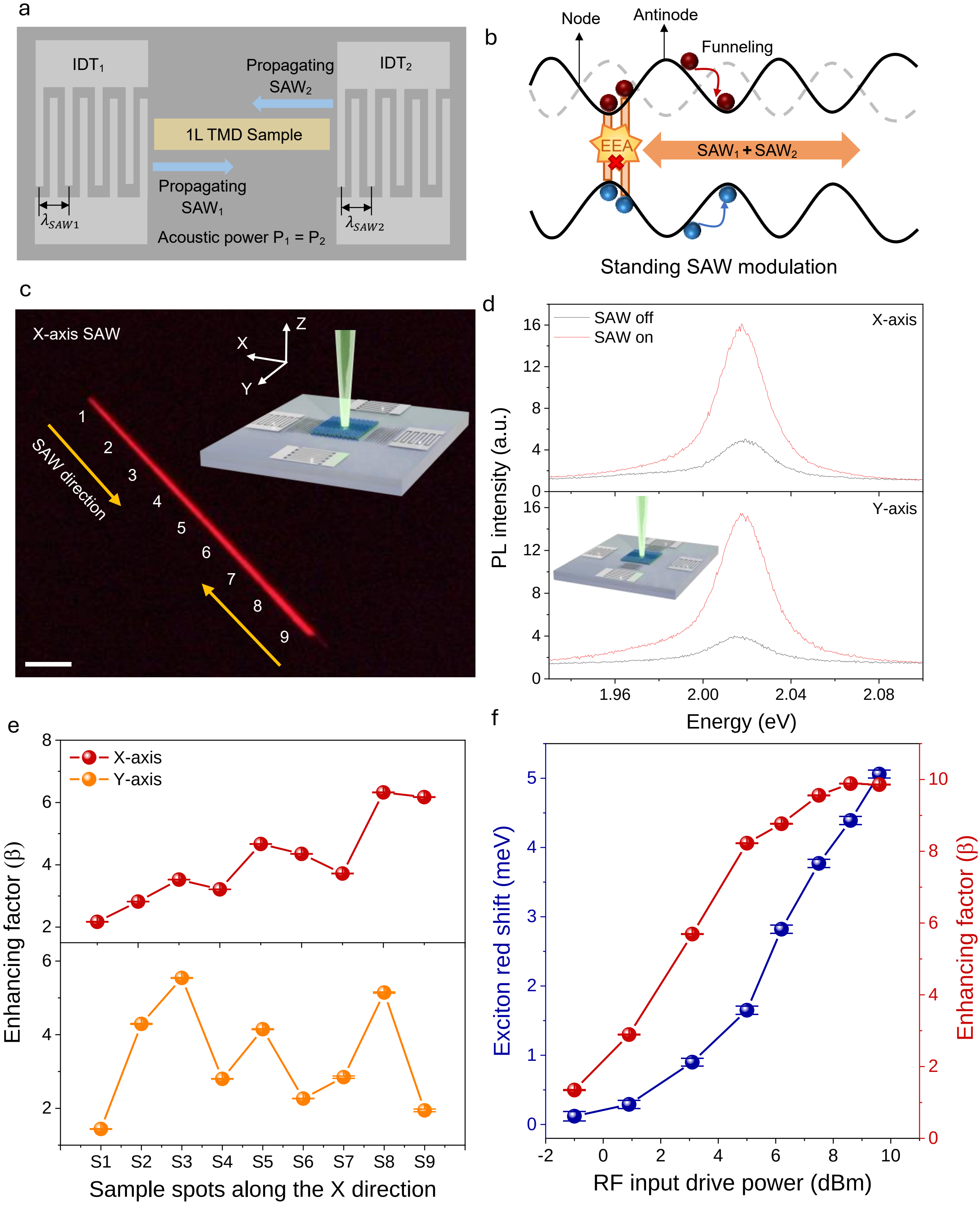

Figure 4

**Figure Captions**

**Figure 1 │ Modulation of exciton emissions in a TMD monolayer at room temperature by a travelling SAW. a**, Schematic illustration of the experimental device with a monolayer TMD located on the SAW delay line. The SAW modulation is launched from IDTs and propagates toward the sample. The IDT excites SAW at 178 MHz with a wavelength of around 22 μm. **b**, PL mapping image of the device shown in the inset, illustrating that the sample, with nearly uniform emission, is aligned with the transport direction of the acoustic wave. The inset shows optical microscope image of a $WS_2$ monolayer stripe sample (outlined by white dotted lines) that is transferred onto a $LiNbO_3$ substrate, positioned close to and perpendicular to the IDTs. The arrow indicates the direction of SAW propagation. S1 and S8 refer to the first and eighth sample spots collected for the measurement conducted in this study. **c**, SAW-induced modulation of the energy bandgap in 2D semiconducting materials, resulting from piezoelectric fields. The blue dashed oval highlights the region of exciton formation, while the black arrows indicate the drift of free carriers, leading to a dissociation of excitons. **d**, Measured spot-dependent PL spectra from a transferred $WS_2$ on the delay line of a SAW device, with SAW on (red) and off (black) at an input power of 10 dBm, taken from locations close to and far away from the IDTs (as shown in the inset) under an excitation power of 2 mW with 532 nm CW laser. The exciton emission appears to be completely quenched at the nearest spot. **e**, Extracted quenching factor (top) and red shift (bottom) in PL peak energies from SAW modulation as a function of positions (averaged as eight spots) along the X direction, corresponding to the wave propagation, compared to the PL emission without SAW. The error bars represent the fitting uncertainty observed across multiple data analyses. **f**, Schematic cross-section diagram showing the modulated monolayer under a propagating non-perfect Rayleigh-SAW, where exciton energy and emission intensity are modified as they transport from the excitation spot to the edge.

**Figure 2 │ Optical visualization of spatial-temporal dynamics in carriers drifting within $WS_2$. a**, Schematic illustration of the experimental setup, featuring a monolayer TMD positioned on the SAW delay line under a photoluminescence microscope. The light source illuminates the entire sample area, resulting in exciton excitation across the entire surface. **b**, Measured real-space PL mapping images of the device captured at different stages of SAW modulation: before SAW, with SAW on at 0, 0.5, 1.0 and 1.5 seconds, respectively. The arrow indicates the propagation direction of the acoustic wave. The scale bar is 10 μm. **c,** Measured real-space PL mapping images of the device captured at different stages after SAW modulation: SAW off at 0, 0.2, 0.4, 1.0, 3.0, 5.0, 20.0 and 60.0 seconds, respectively. The arrow indicates the propagation direction of the acoustic wave. The scale bar is 10 μm. The color scales below the mapping represent the PL intensity variations across the sample, while the insets display the quantitative line profiles of PL intensity along the SAW propagation path.

**Figure 3 │ Optically accessible modulation on the charge carrier transport in $WS_2$. a**, Schematic representations of SAW modulation effects on excitons and free carriers at different stages at SAW on, highlighting the key processes at 0 (i) and 1.5 seconds (ii), corresponding to the experimental mapping images displayed in Fig. 2b (i) and (ii). **b**, Schematic representations of SAW modulation effects on excitons and free carriers at different stages at SAW off, highlighting the key processes at 0 (i), 5 (ii) and 60 seconds (iii), corresponding to the experimental mapping images displayed in Fig. 2c (i), (ii) and (iii). **c,** Extracted quenching (top) and enhancing (bottom) factors of the device at quenched (blue) spot 5 and enhanced (red) spot 8 as a function of SAW-driven power from 0 to 10 dBm. The error bars represent the fitting uncertainty observed across multiple data analyses. **d**, Extracted quenching (left) and enhancing (right) factors of the device at quenched (blue) spot 7 and enhanced (red) spot 8 as a function of laser excitation power from 1 mW to 10 mW. The orange dashed area indicates the regime that is dominated by a different modulation mechanism compared to the blank region with a declining trend. The error bars represent the fitting uncertainty observed across multiple data analyses.

**Figure 4 | Modulated exciton emission and carriers drifting under a standing SAW wave.** **a,** Schematic of the device with a monolayer TMD on the delay line of a standing SAW wave, generated by two IDTs with the same resonant frequency and driving power. The SAW modulations are launched from two IDTs individually and propagate toward each other. **b**, Standing SAW-induced modulation on 2D TMD monolayers, where the SAW effects result from the overlay of two identical propagating waves, leading to a superimposed standing wave resonance pattern with two distinct extreme positions of nodes and antinodes. The orange arrows indicate the opposite directions of two acoustic waves. The red and blue arrows represent the exciton funnelling effect and the red cross highlights the suppressed exciton-exciton annihilation. **c**, PL mapping image of the device with a schematic illustration of the standing SAW, showing the driving direction along the X-axis, parallel to the sample orientation. The scale bar is 10 μm. The distance between each spot is around 8 μm. **d**, Measured PL spectra from the sample illustrated in (**c**) and Fig. S10, showing the effects of a standing SAW on the monolayer sample with both parallel (top) and perpendicular (bottom) applied modulations. **e**, Extracted enhancing factors (β) from measured PL spectra as a function of position along the sample with horizontally (top, along the X-axis) and vertically (bottom, along the Y-axis) applied standing SAW. **f**, Extracted peak shift in PL energies between emissions with and without standing SAW modulation, along with extracted enhancing factors produced from SAW effects, as a function of SAW driving power.

**Supplementary Information for**

# Optical Visualization of Carrier Surfing in 2D Monolayers Driven by Surface Acoustic Waves

*Xueqian Sun, Shuyao Qiu, Hao Qin and Yuerui Lu**

Xueqian Sun, Shuyao Qiu, Hao Qin and Yuerui Lu

School of Engineering, College of Engineering, Computing and Cybernetics, The Australian National University, Canberra, ACT, 2601, Australia

Yuerui Lu

Australian Research Council Centre of Excellence for Quantum Computation and Communication Technology, the Australian National University, Canberra, ACT, 2601 Australia

***** To whom correspondence should be addressed: Yuerui Lu (yuerui.lu@anu.edu.au)

**Supplementary Text 1**

**Exclusion of doping effect from piezoelectric substrate**

The $LiNbO_3$ substrate has been widely studied and is well-known for its high dielectric constant and large piezoelectric effect[1], meaning it can produce electrical charges in response to mechanical pressure or strain[2]. Therefore, to demonstrate effective modulation of ultra-thin monolayers with SAWs, the doping effect from the substrate, along with our experimental observations, has to be excluded.

We conducted experimental measurements on various samples to demonstrate that the doping effect has a negligible contribution to our results.

1. An extra-long sample that exceeds the distance it travels during one period was employed as shown in Figure S13 (a) and (b). The wavelength of SAW modulation is calculated by:

$$\lambda = \frac{\mathrm{v}}{f} \tag{1}$$

The frequency used here is 178MHZ, the velocity of SAW on LiNbO3 is 3990m/s, thus the wavelength is 22 μm. But the length of the sample shown above is around 80 μm, which is much longer than the wavelength. If the doping effect from the substrate dominates the observed modulation, we will be able to see the alternating variations of light and dark patterns with a period corresponding to SAW period.

2. A carked sample with two edges along the wave propagation was employed as shown in Figure S13 (c) and (d).

In the measured mapping images, the two brightest spots appeared at the ends of the two cracked samples when the SAW was turned on. This is consistent with our interpretation of the experimental results discussed in the main text. If the observation were significantly influenced by the doping effect from the substrate, we would expect to see periodic modulation along the wave rather than the emission being concentrated at the edges of the sample.

3. The h-BN half-capsulated samples were employed where the h-BN was inserted between $LiNbO_3$ substrate and $WS_2$ monolayer, as shown in Figure S13 (e) and (f) and Figure S14 (a) and (b).

The layer of h-BN acts as an insulating layer[3] to prevent the charge transfer and doping effect to the top materials from the bottom substrate. However, consistent modulation is still observed in these samples.

**Exclusion of doping effect and confirmation of dominate Type-II modulation**

4. A suspended bubble was employed on the monolayer sample for SAW modulation, as shown in Figure S13 (e), (g) and (h).

The suspension of the bubble region physically separates the material from the substrate, hindering the doping effect and providing additional strain. As shown in the PL spectra in Figure S13 (g) and (h), the SAW modulation on the left side of the bubble and at the top of the bubble with the highest strain indicates a comparable quenching efficiency. This further excludes the role of the doping effect and highlights the dominant influence of Type II modulation in our observations.

**Evidence of transportation of light with SAW**

To further demonstrate the transport of light and carriers in the semiconducting layer, a thick conducting graphene layer was transferred on top of the $WS_2$ monolayer, covering the edge of the sample, as shown in Figure S14 (c). Compared to the sample without graphene (Fig. S14 (a) and (b)), the dissociated carriers and excitons accumulated at the sample edge along the SAW propagation. However, the brightest spot in the $WS_2$/hBN sample was significantly quenched when the graphene layer was present, even with the SAW on (Fig. S14 (d)). The emission was transported away through the conducting graphene, and this is consistent with previous reports of graphene's quenching effect on exciton emission in TMD monolayers[4].

## Supplementary Text 2

### Selection of $LiNbO_3$ substrate

The extensively used piezoelectric substrate of 128° Y-cut $LiNbO_3$ has been chosen to generate SAW modulation from the IDTs through the piezoelectric effect, because of its exceptional properties, including a high piezoelectric coupling constant (~5.36%)[5] and fast SAW travel velocity (3990-3994 m/s)[5-7]. These attributes are crucial for optimizing the performance of SAW resonators and facilitating efficient energy conversion. Combined with the ultra-thin nature of the monolayers and their proximity to the surface waves, they enable exceptionally strong interactions between materials and waves, thereby allowing for an effective investigation of SAW modulation effects on 2D films.

**Supplementary Text 3**

**Evidence of non-perfect Rayleigh SAW modulation along the sample**

From the measured PL spectra of the elongated sample, the trion and exciton contributions were fitted separately (Fig. S5(a)), with their relative ratio indicating the carrier concentration in the material, which facilitates the formation of charged trions. As shown in Figure S5(b), the trion-to-exciton ratio remains almost constant across the sample. However, when the SAW is turned on, the trion contribution increases significantly at the spot closest to the IDTs, while showing a declining trend along the sample as the SAW propagates. At the farthest spot from the transducer, the trion ratio becomes comparable to the unmodulated sample, indicating that the dissociated free carriers decrease dramatically farther from the IDTs. This suggests that the SAW modulation weakens as the wave travels away from the transducer.

**Supplementary Text 4**

**Evidence of exciton dissociation and increased exciton emission at the edge by Time-Resolved Photoluminescence**

TRPL measurements were conducted at the quenching spots (Fig. S6(a)) and the enhancing spots (Fig. S6(b)) near the edge, showing different effects (Fig. S6(c)) due to SAW modulation. The TRPL decay of exciton is faster under the influence of SAW from spot 1 to spot 7, attributed to the ionization of excitons resulting from Type II modulation. In contrast, at spots closer to the sample edge, the emissions are enhanced due to the accumulation of excitons, leading to increased exciton density. The corresponding increase in exciton lifetime at these spots aligns with previous reports, where exciton lifetimes were extended at high exciton densities[8, 9], as non-radiative exciton-exciton annihilation[10] was suppressed by strain.

**Supplementary Text 5**

**Piezoelectric field calculation and free-carriers drift velocity**

Calculating the piezoelectric field generated by a SAW involves understanding the interaction between mechanical strain and electric fields in piezoelectric materials. We focus on lateral electric field here and to simplify the calculation, we assume the conversion efficiency from electrical power to acoustic power is 100% to estimate the maximum piezoelectric field. The surface acoustic wave generates an alternating electric field due to the piezoelectric effect and the surface potential is directly related to the input power of the SAW.

For the SAW device used in our study, SAW wavelength is around 22 μm based on the equation of $\lambda = \frac{\mathrm{V}}{f}$, corresponding to the distance of two electrons and two gaps between electrodes.

The electric field can be described as the potential difference per unit distance between two points:

$$E = -\frac{\nabla \mathrm{V}}{d} \tag{2}$$

Where E is the estimated electric field, ∇V is the electric potential (voltage) difference, and $d$ is the distance over which the potential difference is measured. With an input power of 10 dBm, the corresponding strength of electric field is around 3568.92 V/cm, which agrees with the theoretically predicted strength for exciton dissociation[11, 12] and is consistent with previous reported lateral fields generated in SAW devices[13, 14].

Then the drift velocity of electrons is estimated by:

$$u = \mu \times E \tag{3}$$

where $u$ is drift velocity under the external force, μ is the electron mobility of material, and E is the lateral electric field. We extract a drift velocity of approximately $1.6 \times 10^3$ m/s with intrinsic electron mobility of 44 $cm^2$/Vs at room temperature[15-17], aligning with the previously reported electric-field-dominated transport of electrons in TMD monolayers[18-20]. This demonstrates the unusual transport of carriers but also indicates a weak coupling between the traveling wave and free electrons.

**Ionized free-carriers drift velocity under the photoluminescence everywhere**

However, in the PL mapping conducted in this work, the light source remains continuously on, resulting in the constant generation of excitons under continuous excitation. This contrasts with single, one-time excitation and allows us to observe the slower drift[18] of dissociated carriers at significantly higher carrier concentrations. Here, the exciton generation rate (G) is estimated based on the photon flux and the absorption coefficient of 1L $WS_2$.

The excitation light source we used in this study spans from 480 nm to 500 nm. For a given wavelength (λ), the photon energy is calculated as follows:

$$E = \frac{hc}{\lambda} \quad (4)$$

where $h$ is Planck's constant, and $c$ is the speed of light. Thus, the photon energy in our study ranges from $3.97\times10^{-19}$ J (for 500 nm) to $4.14\times10^{-19}$ J (for 480 nm).

The photon flux is defined as the number of photons incident per second per unit area and is expressed by:

$$\phi = \frac{P}{A\times E} \quad (5)$$

where P is the incident light power, A is the area of sample and $E$ is the photon energy corresponding to a given wavelength. The photon flux varies across the wavelength range, requiring integration over the range to estimate the total photon flux. The exciton generation rate G is estimated to be proportional to the absorbed photon flux and the absorption coefficient $\alpha(\lambda)$ of the $WS_2$ monolayer, which can be written as:

$$G = \int_{\lambda_2}^{\lambda_1} \phi(\lambda)\, \alpha(\lambda)\, d\lambda \quad (6)$$

For simplification, the exciton generation rate can be estimated by taking the average photon energy and absorption coefficient across the wavelength range, and then multiplying by the total incident power as below:

$$G \approx \frac{P\times\alpha_{Average}}{E_{Average}} = \phi \times \alpha \quad (7)$$

Here, we used the average absorption coefficient $(\alpha)$[21] for the light source, covering from 480 nm to 500 nm, estimated to be $1.3259\times10^{5}$ $cm^{-1}$. The measured power for 480 nm, 490 nm and 500 nm was 35.56µW, 33.38µW and 32.31µW, respectively and the area of incident light in this study is around $1.23\times10^{-10}$ $m^2$. Consequently, the photon flux was calculated as $6.984\times10^{23}$ (480 nm), $6.683\times10^{23}$/s (490 nm) and $6.618\times10^{23}$/s (500 nm). The corresponding exciton generation rates were estimated as $9.259\times10^{30}/m^3/s$, $8.861\times10^{30}/m^3/s$ and $8.775\times10^{30}/m^3/s$. Therefore, the total exciton generation rate is approximately $2.691\times10^{31}/m^3/s$. It is known that the electron density of a copper wire is around $8.5\times10^{28}$ $m^3$ and the drift velocity of electrons under an electric field is approximately in the order of $10^{-5}$ m/s. This aligns with our observed drift velocity of dissociated carriers (~16.4 µm/s) at high carrier concentrations. The clear observation of this transport through the image sensor can be understood as the difference between electrical signals and mobile electrons in a wire[22]. The transition speed of an electrical signal can approach a significant fraction of the speed of light, but the electrons themselves move much more slowly inside.

**Supplementary Text 6**

**Time period estimation of the recovery process**

The video and images captured during measurement show that our $WS_2$ sample appeared to recover after the SAW was turned off at 60 seconds (Fig. 2c(iii)). However, closer examination of images taken a few more minutes later reveals subtle differences and additional recovery after SAW deactivation at 60 seconds. As shown in Figure S8, white dotted lines highlight the further recovery region, indicated by a brighter and more uniform left edge at 3 minutes after SAW-off compared to 80 seconds. This suggests that the sample actually requires more than 60 seconds to fully recover from SAW modulation.

**Supplementary Text 7**

**Definition of enhancing factor**

We defined another parameter, β, at the enhancement spots (close to the sample edge and far away from IDTs) to intuitively reflect the difference between the quenching and enhancing regions. β is defined as the intensity ratio of the integrated PL intensities of monolayer with SAW modulation to that of the unmodulated sample:

$$\beta = I_{WS_{2\,SAW\,on}} / I_{WS_{2\,saw\,off}}$$

**Supplementary Text 8**

**Small strain improved quantum yield**

Strain-induced bandgap modulation has been widely studied and typically leads to the quenching of exciton emission in TMD monolayers. However, when mechanical strain is applied at levels below 0.5%, emission can be enhanced due to improved quantum yield—potentially increasing by an order of magnitude—resulting from reduced exciton-exciton nonradiative interactions[10, 23]. In our study with standing waves, the maximum peak shift in exciton resonance is approximately 5 meV, corresponding to around 0.11% strain[24], well under 0.5%, which explains the enhanced emissions observed in our sample despite the presence of mechanical strains.

## Supplementary Text 9

### Fluctuations in enhancing factors under a standing SAW

As shown in Figure 4e, the enhancement factors under the modulation of standing SAW exhibit periodic fluctuations, attributed to the presence of pressure nodes and antinodes. The sample, approximately 68 μm in length (Fig. 4e), exhibits three periods of fluctuation, aligning well with our expectations and corresponding to the SAW wavelength of 22 μm. Notably, the X-axis SAW applied along the sample orientation shows a slight increase in enhancement along the sample (top), while the Y-axis SAW does not show any significant changes (bottom). This difference is attributed to the sample not being perfectly centred between the two horizontal IDTs, resulting in varying modulation strength along the sample, as we illustrated earlier. For the vertically applied SAW, which is perpendicular to the sample orientation, the distance from each point on the sample to the vertical IDTs remains constant, leading to a uniform modulation effect without pronounced responses at specific locations.

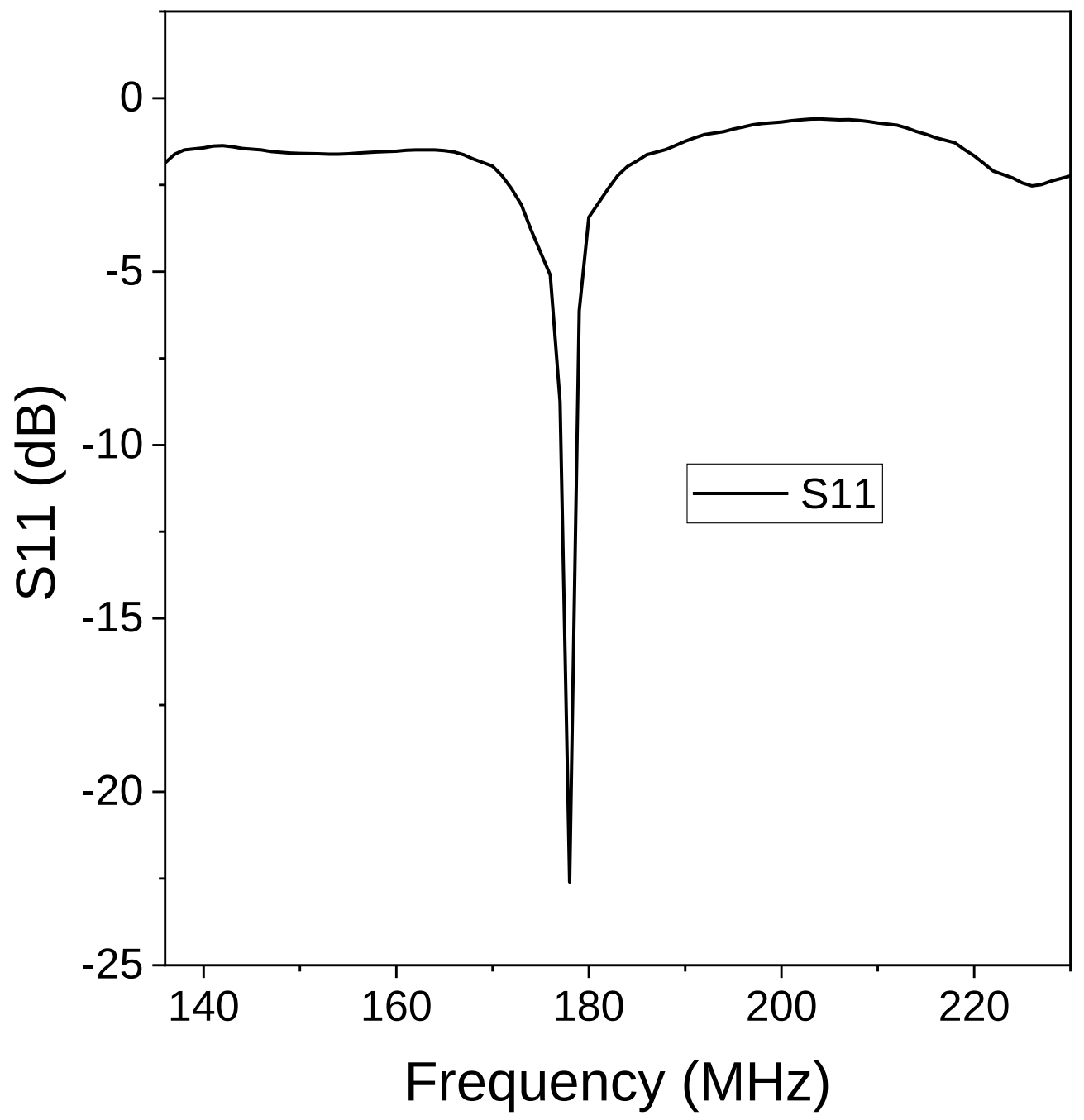

**Supplementary Fig. 1 |** Meaured electric reflection co-efficient (S11) of the SAW resonator used in the experiments at room temperature.

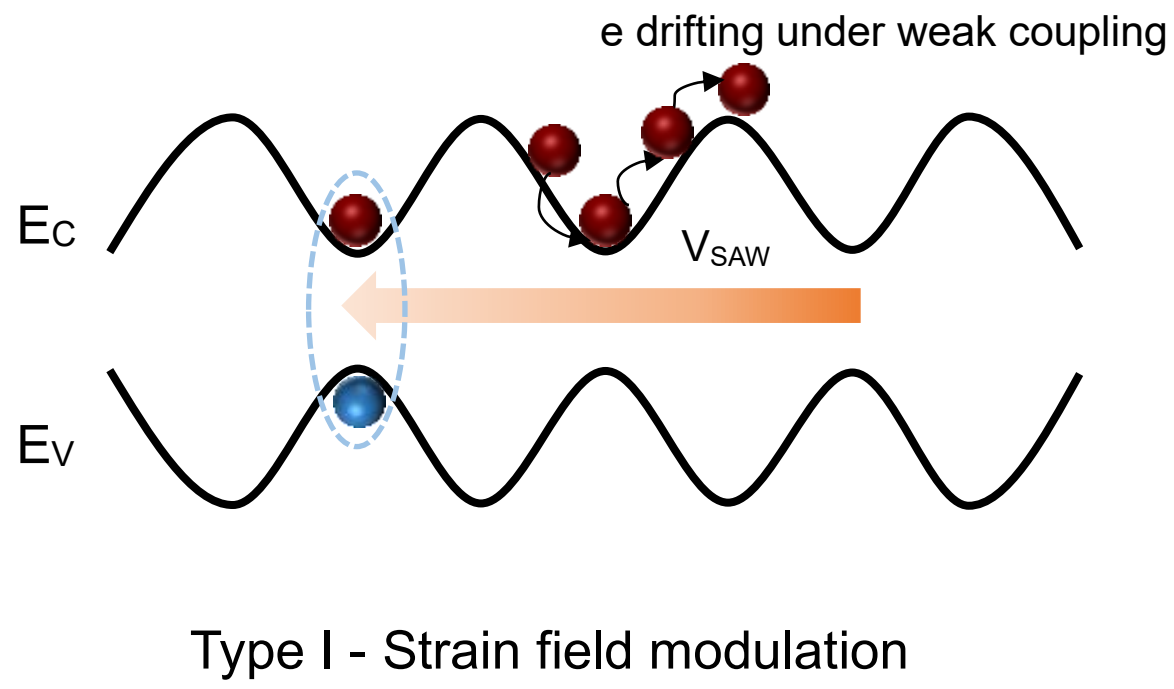

**Supplementary Fig. 2 |** SAW-induced modulation of the energy bandgap in 2D semiconducting materials, resulting from mechanical strain. The blue dashed oval highlights the region of exciton formation, while the black arrows indicate the drift of free carriers under a weak coupling with acoustic waves.

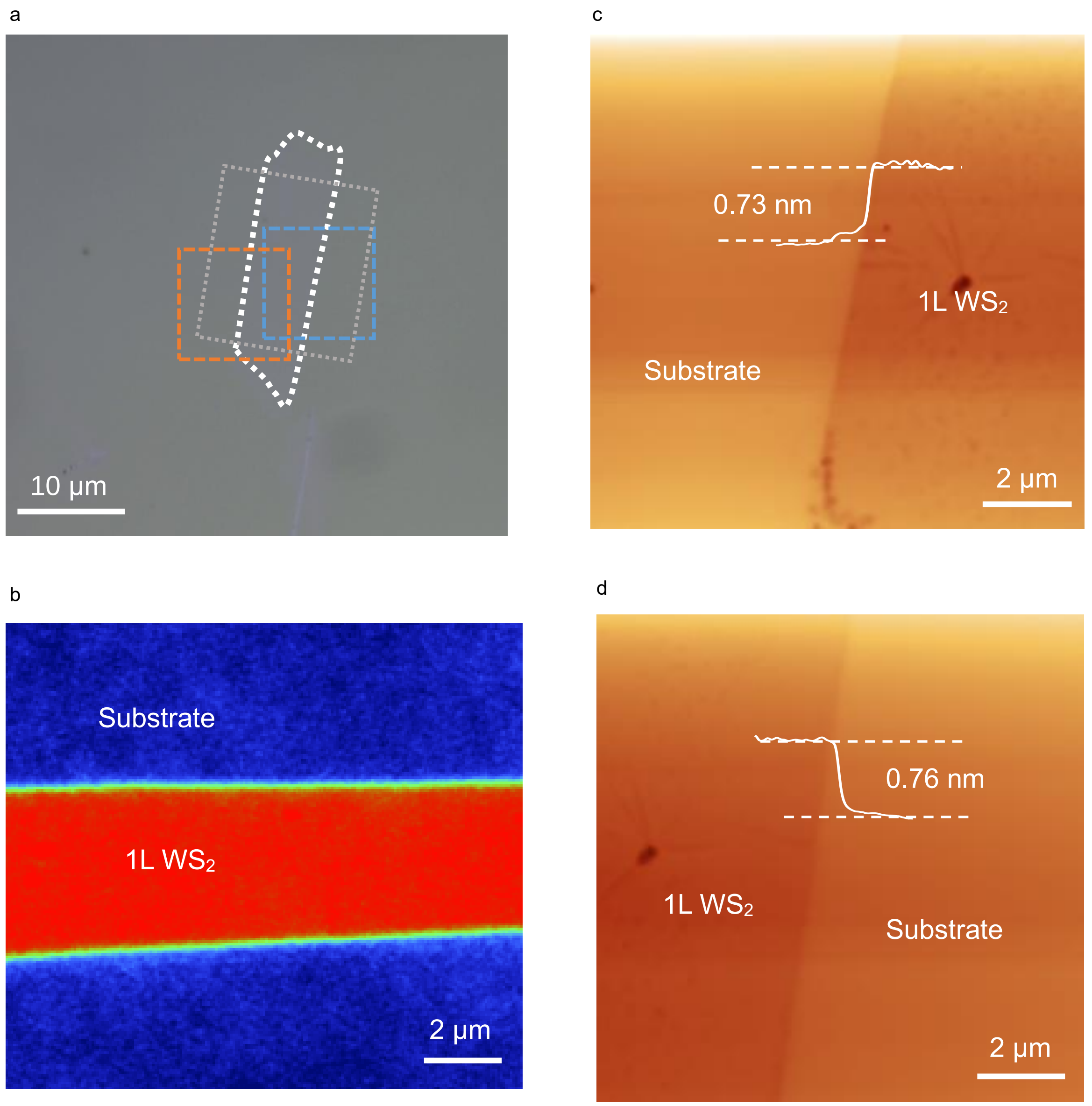

**Supplementary Fig. 3 | a**, Optical microscope image of the $WS_2$ monolayer (outlined by white dotted line) on the transparent $LiNbO_3$ substrate. **b,** Phase shifting interferometry (PSI) image of the region inside the grey box indicated by the dashed line in (**a**), showing the uniform optical thickness[25] of the monolayer sample. **c-d**, AFM images of the area in (**a**) indicated by the orange and blue dashed lines. The inset shows a height profile taken along the dashed line in (**c**) and (**d**), confirming the presence of monolayer $WS_2$.

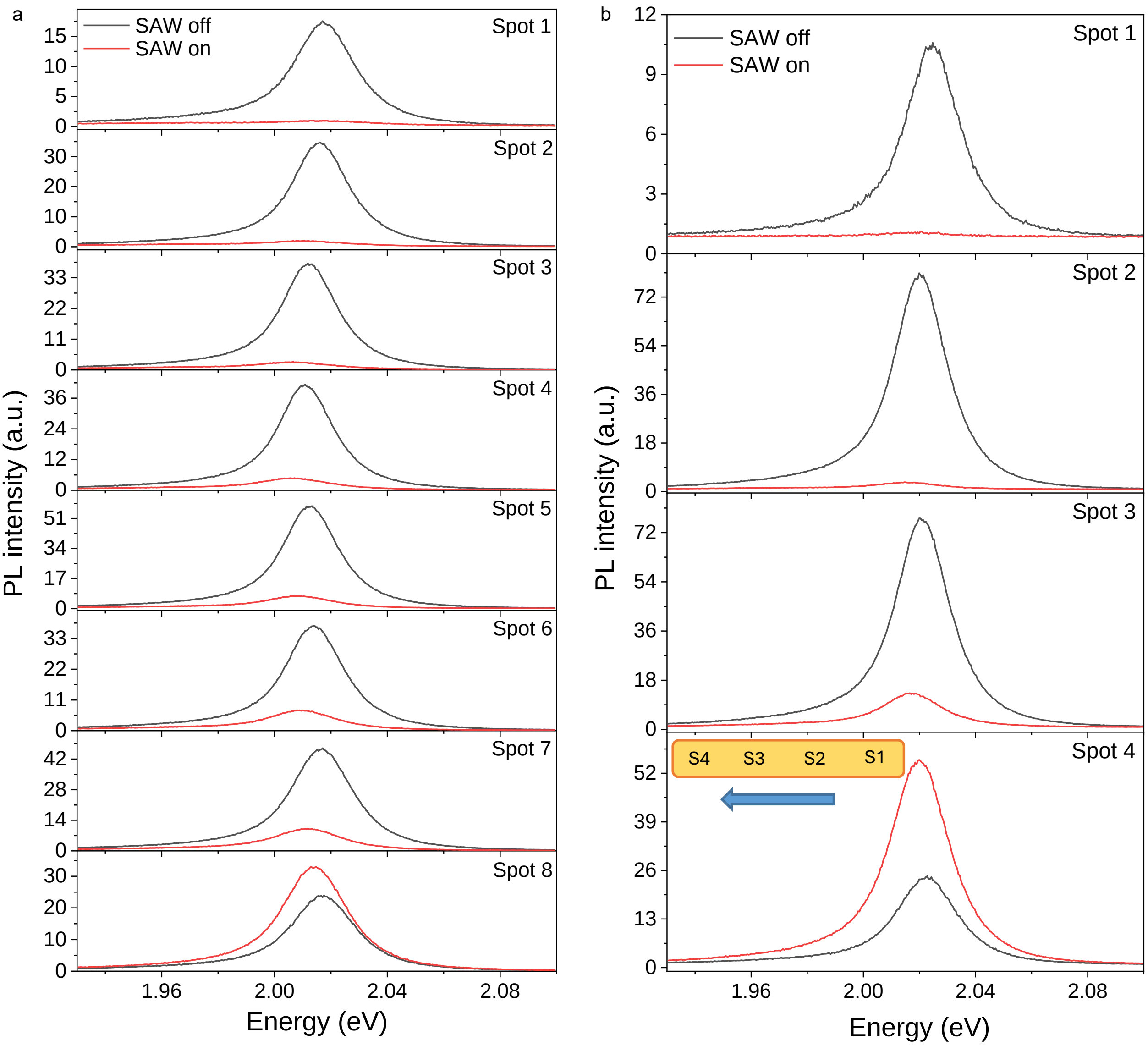

**Supplementary Fig. 4 | Measured spot-dependent PL spectra along the SAW propagation at room temperature. a-b**, Measured PL spectra from various positions along the samples with the propagation direction of SAW (blue arrow). The samples were divided into eight points (sample 1) (**a**) and four points (sample 2) (**b**), respectively.

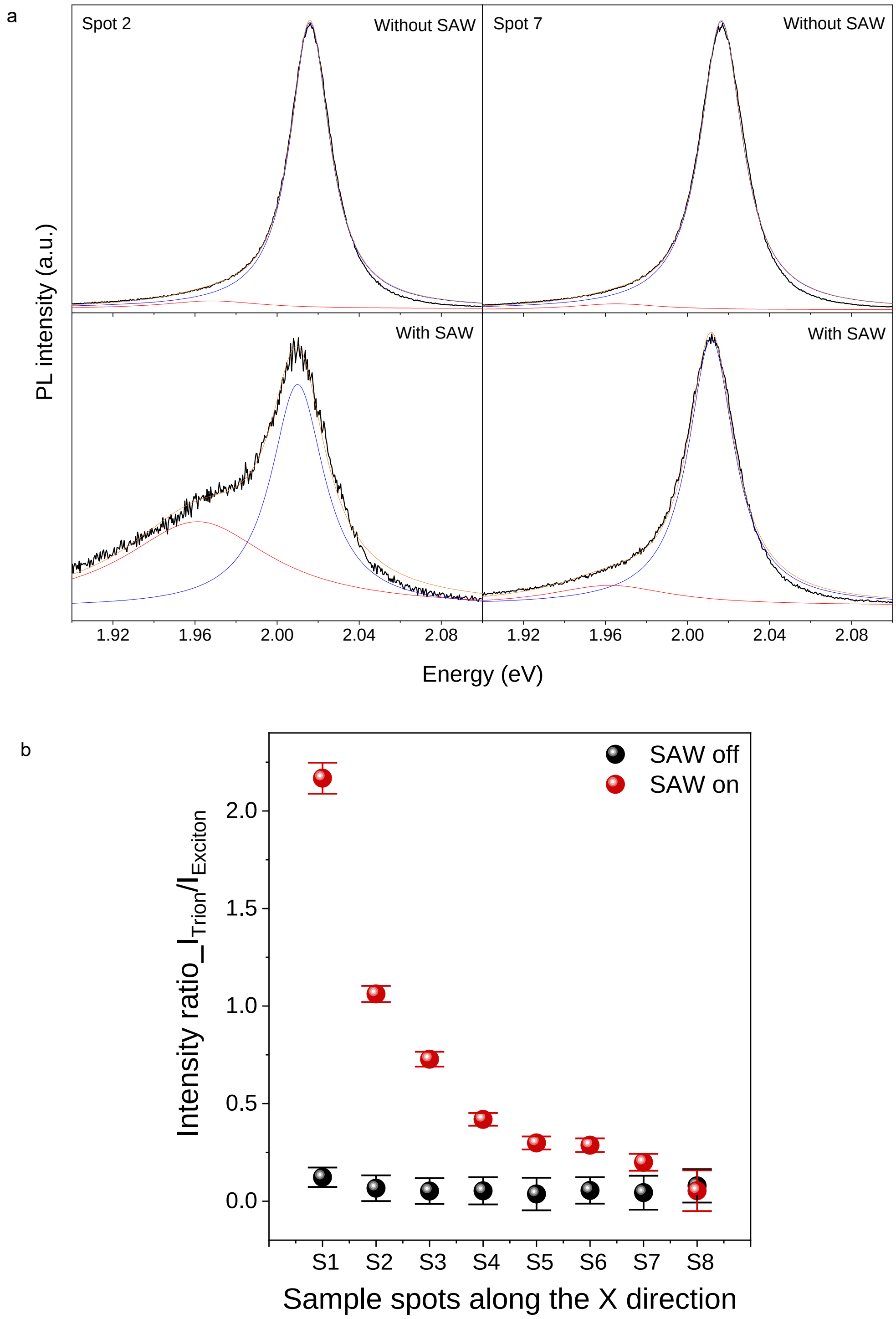

**Supplementary Fig. 5 | Extracted spot-dependent trion contribution as a function of position. a**, Comparison of trion (red curve) and exciton (blue curve) of monolayer $WS_2$ at Spot 2 (left) and 7 (right) with (bottom panel) and without (top panel) SAW modulation. The spectrum was fitted with two Lorentz peaks. **b**, Extracted peak ratio of Trion to Exciton as a function of position from PL spectra measured with (red) and without (black) SAW modulation.

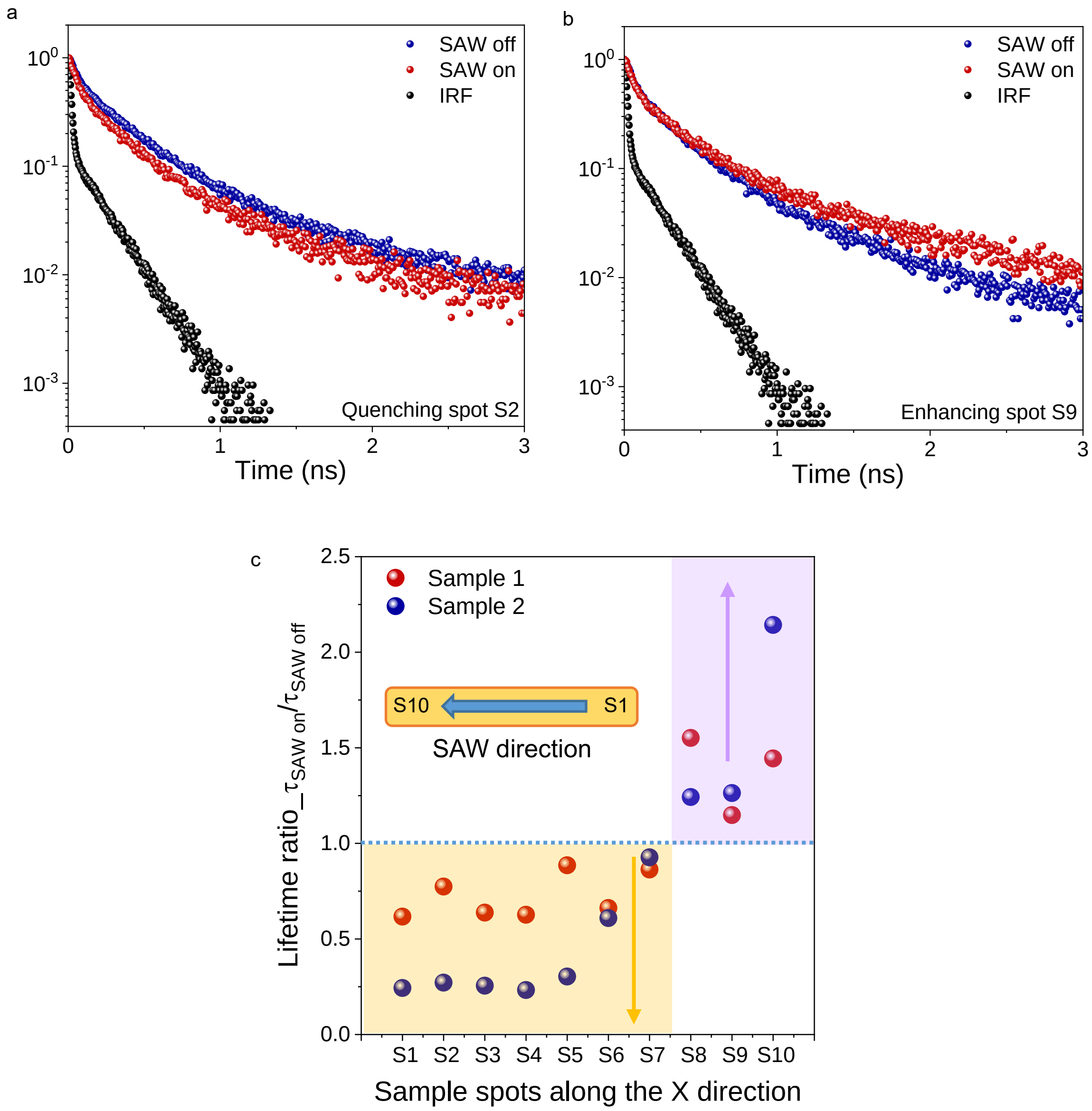

**Supplementary Fig. 6 | Time-resolved photoluminescence of $WS_2$ monolayer under the SAW driven modulation. a-b**, Measured local time-resolved PL emission (normalized) from $WS_2$ monolayer with (red) and without (blue) SAW modulation at quenching spot 2 (**a**) and enhancing spot 9 (**b**) respectively. The decay curve was fitted by deconvoluting the data from the instrument response function (IRF), yielding a lifetime of 473 ps, 425 ps, 444 ps, and 565 ps, for exciton at quenching location without SAW (blue in left panel), with SAW (red in left panel), at enhancing location without SAW (blue in right panel), with SAW (red in right panel), respectively. **c**, Extracted lifetime ratio from measured decay curves of $WS_2$ monolayers with SAW on to the decay curve with SAW modulation off as a function of position. The yellow and purple arrows indicate reduced and improved lifetimes with SAW on, respectively.

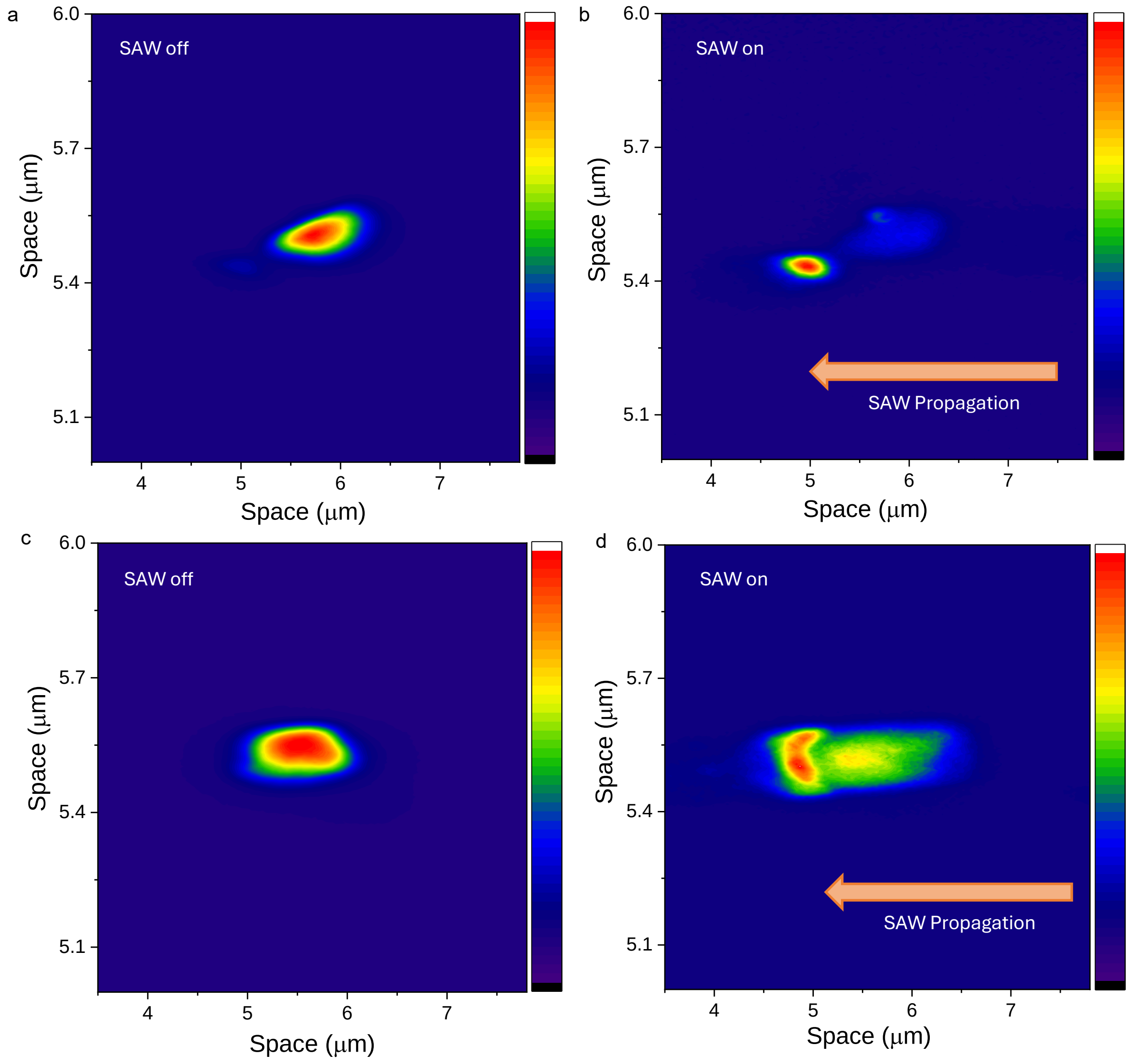

**Supplementary Fig. 7 | a-d**, Real-space PL mapping with a single spot excitation when SAW is off (**a**, **c**) and on (**b**, **d**) with 2mW excitation power for sample 1 and 2. The emission spot away from the laser excitation along the SAW propagation and the quenching of local PL indicate a remote emission and proven that the excitons are ionized and transported with SAW.

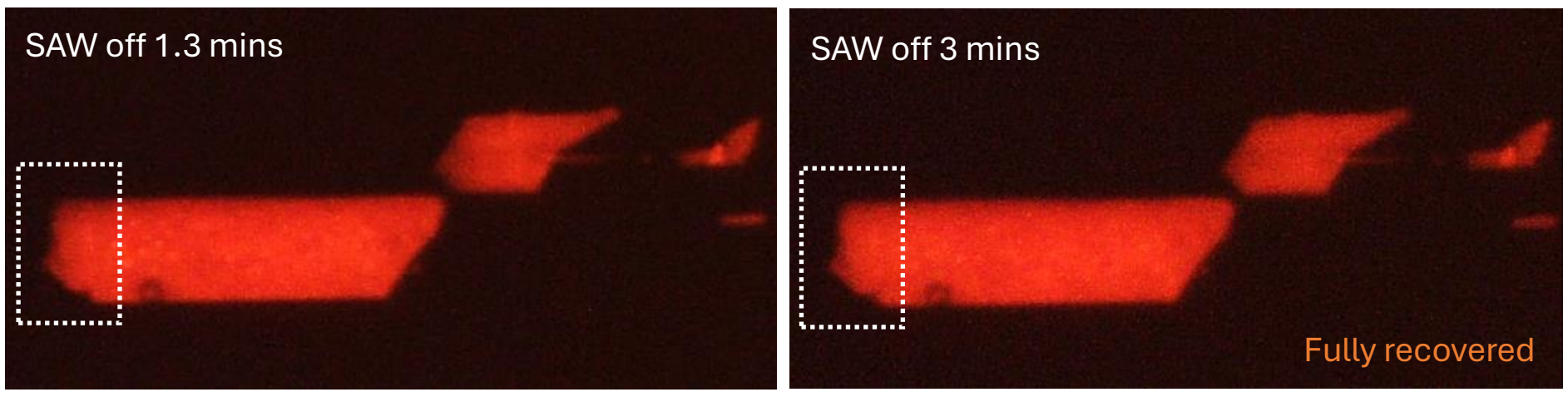

**Supplementary Fig. 8 |** PL mapping images of the device shown in Figure 2 captured at saw off after 1.3 minutes and 3 minutes respectively.

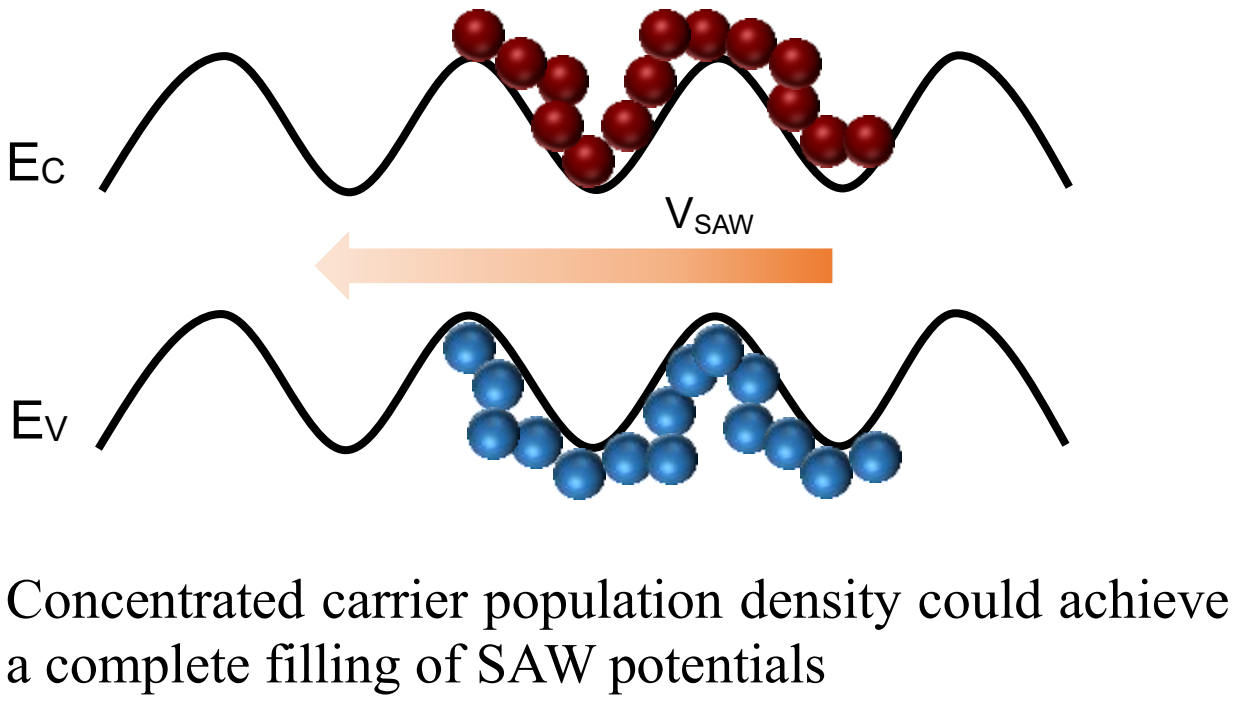

**Supplementary Fig. 9 |** Schematic illustration of the SAW modulation on the monolayer TMD when the exciton density is higher with a high excitation pumping power. When the SAW potential wells are completely filled or overloaded, the SAW effects become inefficient, and the related experimental phenomenon could show a saturation indicator.

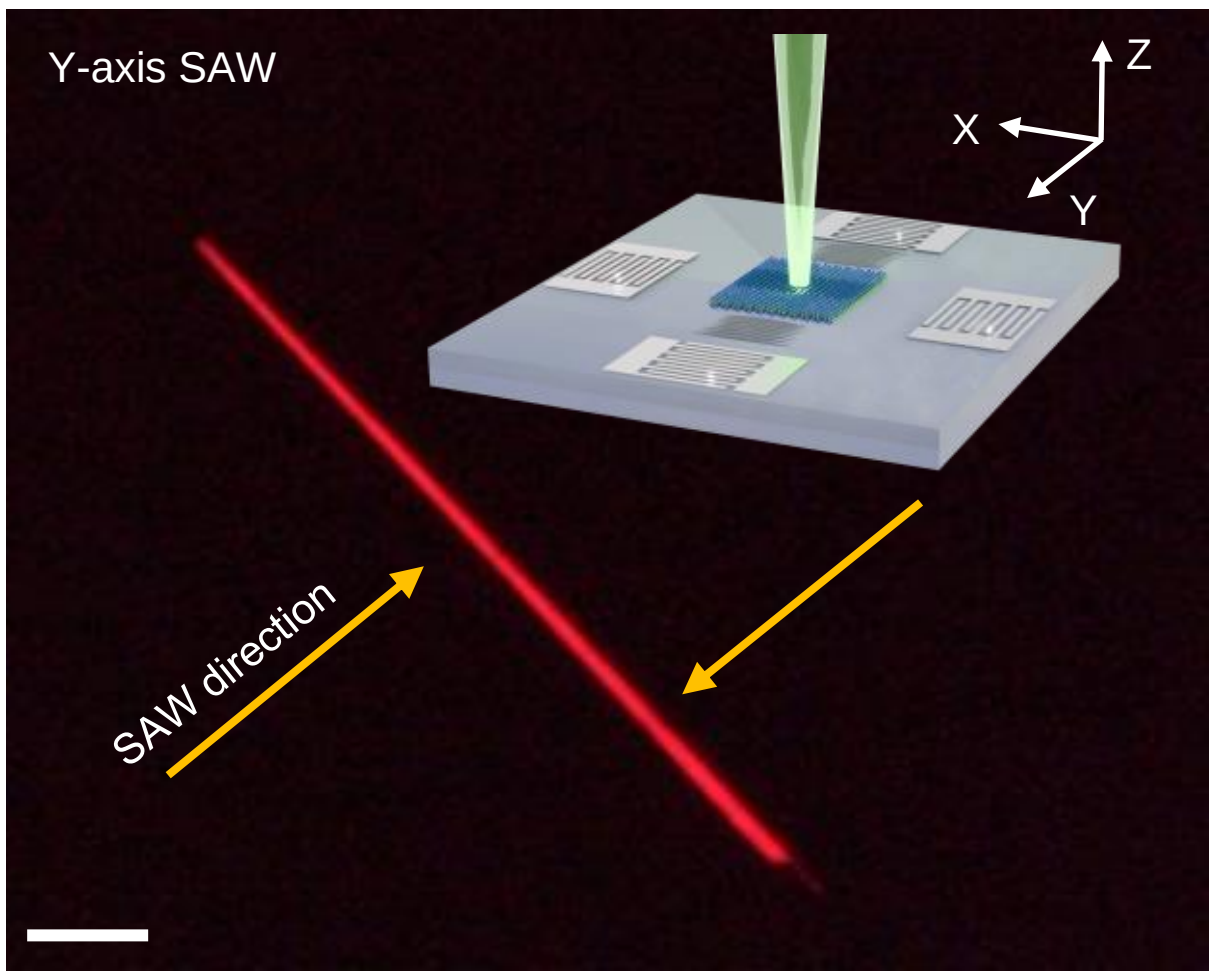

**Supplementary Fig. 10 |** PL mapping image of the device with a schematic illustration of the standing SAW, indicating the driving direction perpendicular (along the Y-axis) to the sample orientation. The scale bar is 10 μm.

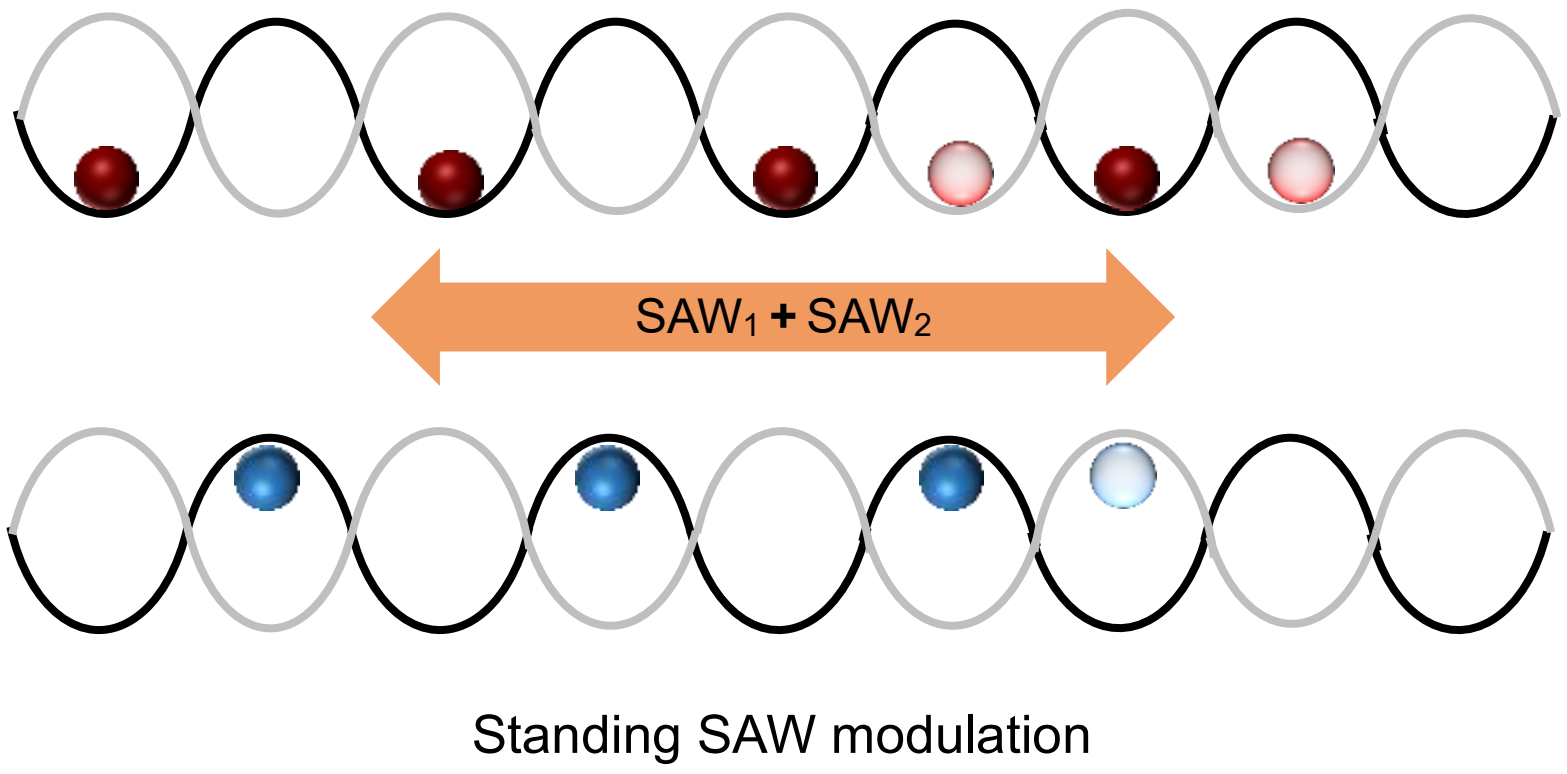

**Supplementary Fig. 11 |** Standing SAW-induced modulation with two opposite travelling SAW with identical wave patterns. The orange arrows indicate the opposite directions of two acoustic waves, the dark and light spheres (for both red and blue) represent electrons and holes travelling with the two different SAWs, respectively.

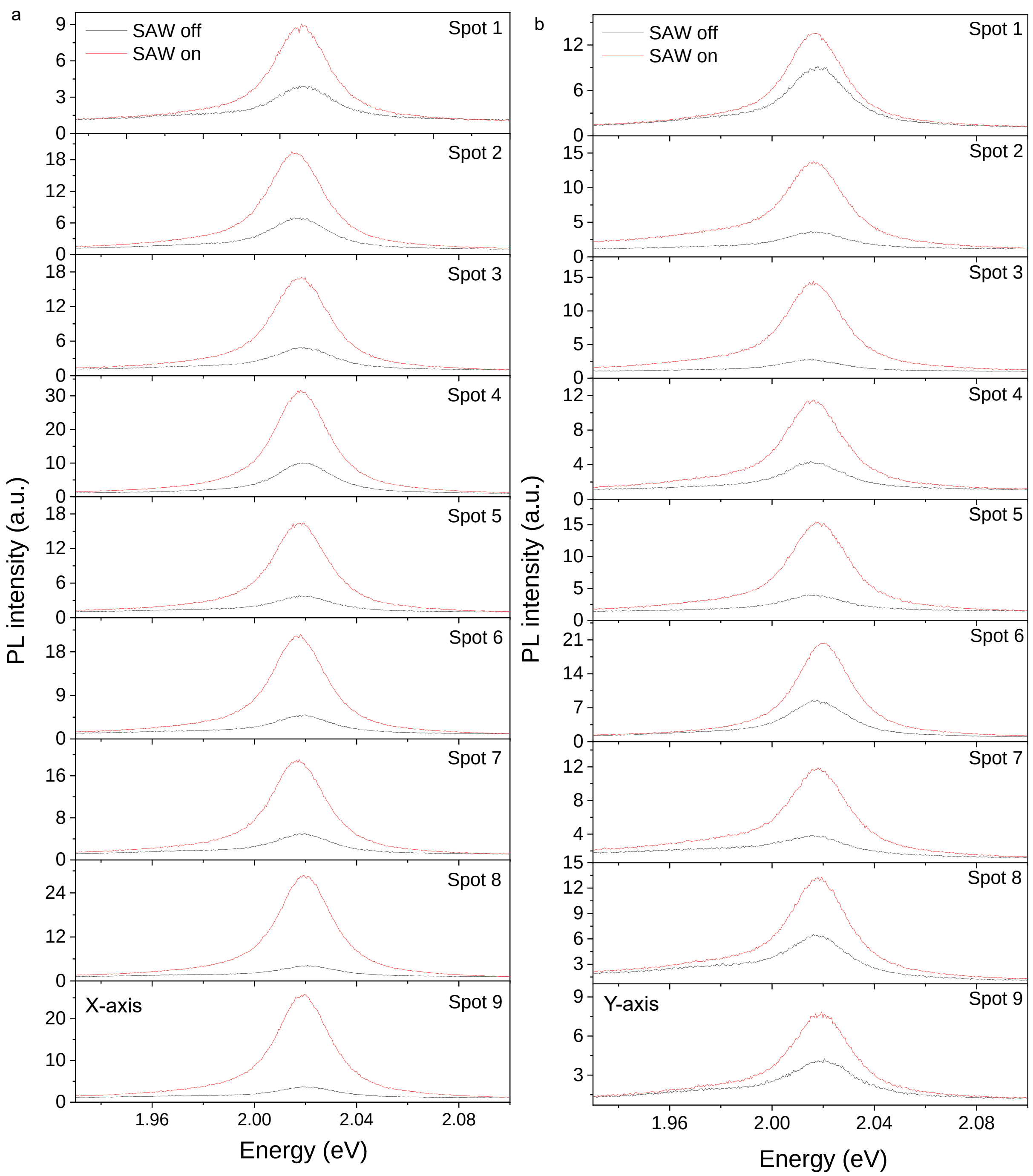

**Supplementary Fig. 12 | Measured spot-dependent PL spectra along the sample with standing SAWs at room temperature. a**, Measured PL spectra from various positions along the sample with the SAW propagating parallel (X-axis) to the sample (Fig. 4c). **b**, Measured PL spectra from various positions along the sample with the SAW propagating perpendicular (Y-axis) to the sample (Fig. S10).

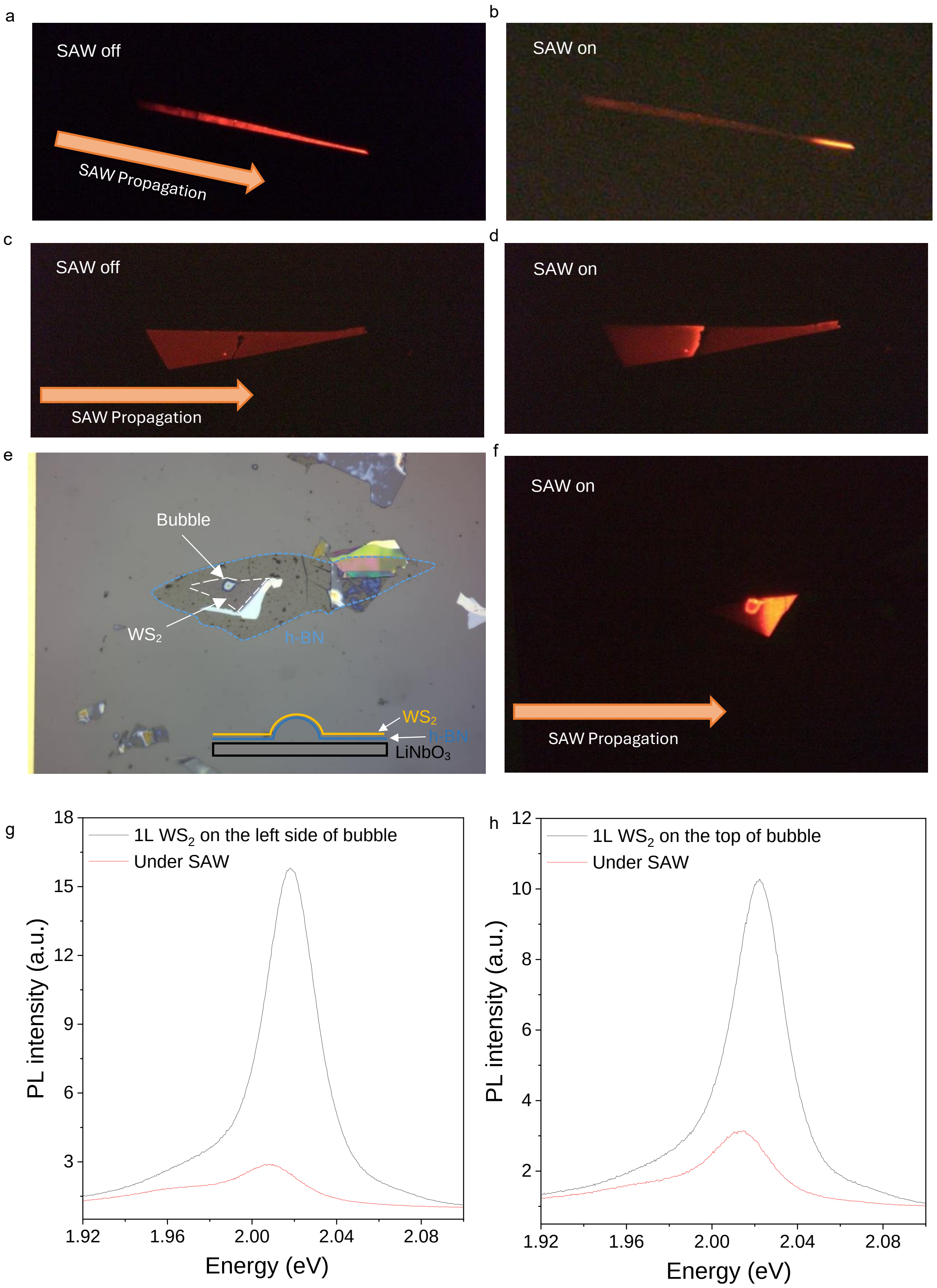

**Supplementary Fig. 13 | Exclusion of doping effect from piezoelectric substrate. a-b**, Real-space PL mapping images of the extra-long SAW device with SAW modulation OFF (**a**) and ON (**b**), propagating along the sample. **c-d**, Real-space PL mapping images of the cracked SAW device with SAW modulation OFF (**c**) and ON (**d**), propagating along the sample. **e**,

Optical image of the SAW device constructed by $WS_2$/hBN/$LiNbO_3$ with a nano-bubble on the hBN insulating layer. **f**, Measured PL mapping images of the device illustrated in panel (**e**) with SAW ON, propagating along the sample. **g-h**, Measured PL spectra from the spot on the left side (**g**) and top (**h**) of bubble with (red) and without (black) SAW modulation from left side.

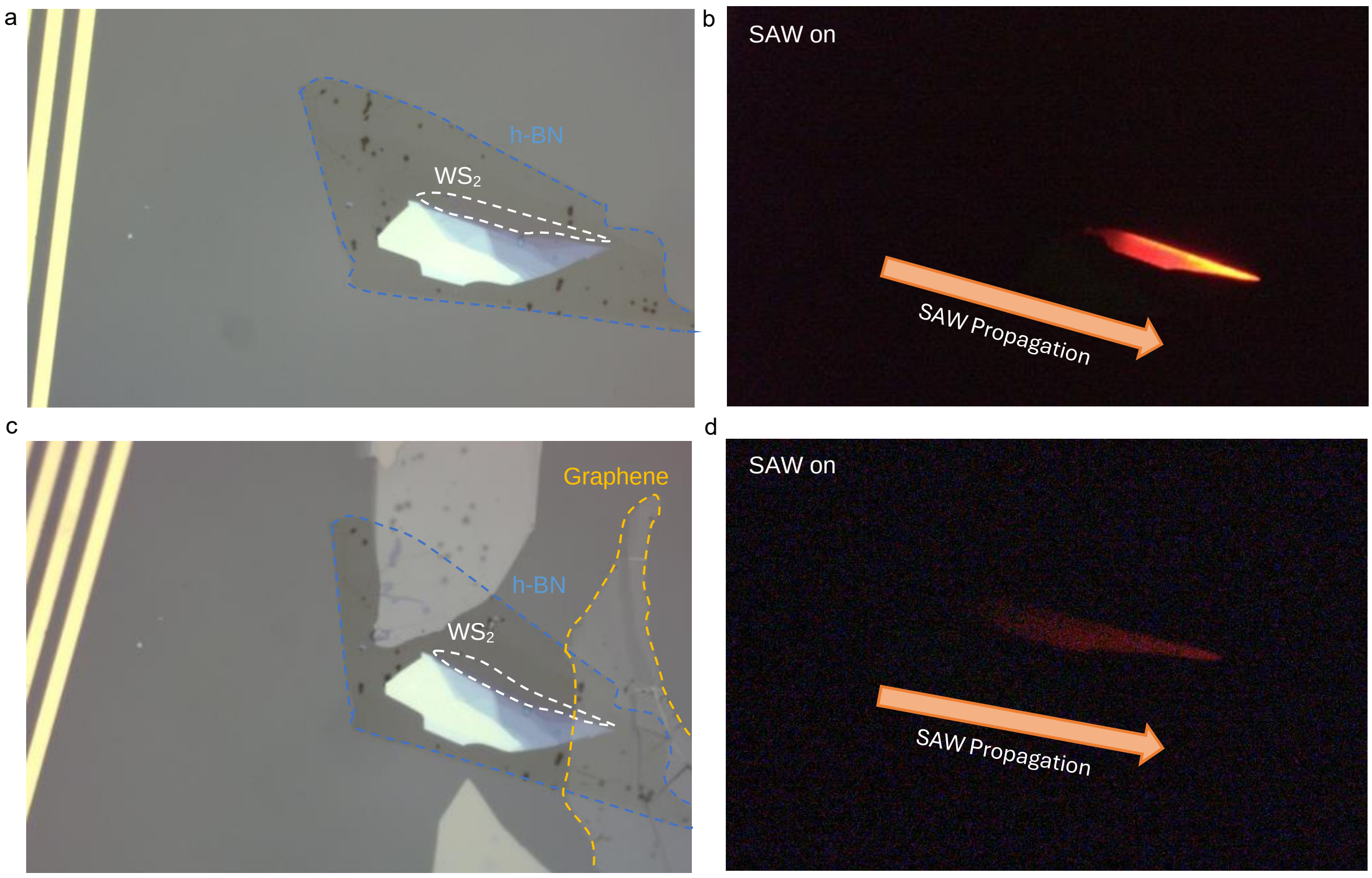

**Supplementary Fig. 14 | Further validation of carrier and light transportation. a**, Optical image of the SAW device constructed by $WS_2$/hBN/$LiNbO_3$. **b**, Real-space PL mapping images of the device displayed in (**a**) with SAW modulation ON and propagating along the sample. **c**, Optical image of the SAW device constructed by Graphene/$WS_2$/hBN/$LiNbO_3$ where the graphene is stacked to only cover the end of device fabricated in (**a**). **d**, Real-space PL mapping images of the device displayed in (**c**) with SAW modulation on and propagating along the sample.

**Supplementary Movie 1:** Travelling SAW-induced modulation in elongated $WS_2$ monolayer, highlighting two in-plane transitions from right to left with the propagation of carriers.

**Supplementary Movie 2:** SAW-induced modulation in $WS_2$ monolayer under a standing wave, highlighting the periodic fluctuations in the emission.

## References

1. Yue, W. and J. Yi-jian, *Crystal orientation dependence of piezoelectric properties in LiNbO3 and LiTaO3.* Optical Materials, 2003. **23**(1): p. 403-408.
2. Wang, A.C., C. Wu, D. Pisignano, Z.L. Wang, and L. Persano, *Polymer nanogenerators: opportunities and challenges for large-scale applications.* Journal of Applied Polymer Science, 2018. **135**(24): p. 45674.
3. Xu, W., D. Kozawa, Y. Liu, Y. Sheng, K. Wei, V.B. Koman, S. Wang, X. Wang, T. Jiang, and M.S. Strano, *Determining the optimized interlayer separation distance in vertical stacked 2D WS2: hBN: MoS2 heterostructures for exciton energy transfer.* Small, 2018. **14**(13): p. 1703727.
4. Lorchat, E., L.E.P. López, C. Robert, D. Lagarde, G. Froehlicher, T. Taniguchi, K. Watanabe, X. Marie, and S. Berciaud, *Filtering the photoluminescence spectra of atomically thin semiconductors with graphene.* Nature Nanotechnology, 2020. **15**(4): p. 283-288.
5. Ro, R., R. Lee, S. Wu, Z.-X. Lin, and M.-S. Lee, *Propagation characteristics of surface acoustic waves in AlN/128 Y–X LiNbO3 structures.* Japanese Journal of Applied Physics, 2009. **48**(4R): p. 041406.
6. Yang, Y., C. Dejous, and H. Hallil, *Trends and Applications of Surface and Bulk Acoustic Wave Devices: A Review.* Micromachines, 2023. **14**(1): p. 43.
7. Nie, X., X. Wu, Y. Wang, S. Ban, Z. Lei, J. Yi, Y. Liu, and Y. Liu, *Surface acoustic wave induced phenomena in two-dimensional materials.* Nanoscale Horizons, 2023. **8**(2): p. 158-175.
8. Tran, T.T., Y. Lee, S. Roy, T.U. Tran, Y. Kim, T. Taniguchi, K. Watanabe, M.V. Milošević, S.C. Lim, A. Chaves, J.I. Jang, and J. Kim, *Synergetic Enhancement of Quantum Yield and Exciton Lifetime of Monolayer WS2 by Proximal Metal Plate and Negative Electric Bias.* ACS Nano, 2024. **18**(1): p. 220-228.
9. H L, P., P. Mondal, A. Bid, and J.K. Basu, *Electrical and Chemical Tuning of Exciton Lifetime in Monolayer MoS2 for Field-Effect Transistors.* ACS Applied Nano Materials, 2020. **3**(1): p. 641-647.
10. Kim, H., S.Z. Uddin, N. Higashitarumizu, E. Rabani, and A. Javey, *Inhibited nonradiative decay at all exciton densities in monolayer semiconductors.* Science, 2021. **373**(6553): p. 448-452.
11. Miller, D.A.B., D.S. Chemla, T.C. Damen, A.C. Gossard, W. Wiegmann, T.H. Wood, and C.A. Burrus, *Electric field dependence of optical absorption near the band gap of quantum-well structures.* Physical Review B, 1985. **32**(2): p. 1043-1060.
12. Massicotte, M., F. Vialla, P. Schmidt, M.B. Lundeberg, S. Latini, S. Haastrup, M. Danovich, D. Davydovskaya, K. Watanabe, T. Taniguchi, V.I. Fal'ko, K.S. Thygesen, T.G. Pedersen, and F.H.L. Koppens, *Dissociation of two-dimensional excitons in monolayer WSe2.* Nature Communications, 2018. **9**(1): p. 1633.
13. Rocke, C., S. Zimmermann, A. Wixforth, J.P. Kotthaus, G. Böhm, and G. Weimann, *Acoustically driven storage of light in a quantum well.* Physical Review Letters, 1997.

**78**(21): p. 4099.
14. Gulyaev, D.V. and K.S. Zhuravlev, *Mechanisms of exciton photoluminescence quenching in the electric field of a standing surface acoustic wave.* International Journal of Modern Physics B, 2019. **33**(06): p. 1950032.
15. Li, S.-L., K. Tsukagoshi, E. Orgiu, and P. Samorì, *Charge transport and mobility engineering in two-dimensional transition metal chalcogenide semiconductors.* Chemical Society Reviews, 2016. **45**(1): p. 118-151.
16. Cui, Y., R. Xin, Z. Yu, Y. Pan, Z.Y. Ong, X. Wei, J. Wang, H. Nan, Z. Ni, and Y. Wu, *High-performance monolayer WS2 field-effect transistors on high-κ dielectrics.* Advanced Materials, 2015. **27**(35): p. 5230-5234.
17. Ovchinnikov, D., A. Allain, Y.-S. Huang, D. Dumcenco, and A. Kis, *Electrical transport properties of single-layer WS2.* ACS nano, 2014. **8**(8): p. 8174-8181.
18. Iglesias, J.M., A. Nardone, R. Rengel, K. Kalna, M.J. Martín, and E. Pascual, *Carrier mobility and high-field velocity in 2D transition metal dichalcogenides: degeneracy and screening.* 2D Materials, 2023. **10**(2): p. 025011.
19. Jin, Z., X. Li, J.T. Mullen, and K.W. Kim, *Intrinsic transport properties of electrons and holes in monolayer transition-metal dichalcogenides.* Physical Review B, 2014. **90**(4): p. 045422.
20. Li, X., J.T. Mullen, Z. Jin, K.M. Borysenko, M. Buongiorno Nardelli, and K.W. Kim, *Intrinsic electrical transport properties of monolayer silicene and MoS 2 from first principles.* Physical Review B—Condensed Matter and Materials Physics, 2013. **87**(11): p. 115418.
21. Kim, T.J., V.L. Le, H.T. Nguyen, X.A. Nguyen, and Y.D. Kim, *Modeling of the Optical Properties of Monolayer WS2.* Journal of the Korean Physical Society, 2020. **77**(4): p. 298-302.
22. *Model of Conduction in Metals.*
23. Lee, Y., J.D.a.S. Forte, A. Chaves, A. Kumar, T.T. Tran, Y. Kim, S. Roy, T. Taniguchi, K. Watanabe, A. Chernikov, J.I. Jang, T. Low, and J. Kim, *Boosting quantum yields in two-dimensional semiconductors via proximal metal plates.* Nature Communications, 2021. **12**(1): p. 7095.
24. He, X., H. Li, Z. Zhu, Z. Dai, Y. Yang, P. Yang, Q. Zhang, P. Li, U. Schwingenschlogl, and X. Zhang, *Strain engineering in monolayer WS2, MoS2, and the WS2/MoS2 heterostructure.* Applied Physics Letters, 2016. **109**(17).
25. Yang, J., Z. Wang, F. Wang, R. Xu, J. Tao, S. Zhang, Q. Qin, B. Luther-Davies, C. Jagadish, Z. Yu, and Y. Lu, *Atomically thin optical lenses and gratings.* Light: Science & Applications, 2016. **5**(3): p. e16046-e16046.